\documentclass[article,twocolumn,floatfix]{revtex4}

\usepackage{graphicx} 
\usepackage{subfig}
\usepackage{color}

\newcommand{\ket}[1]{\left| #1 \right\rangle}
\newcommand{\bra}[1]{\left\langle #1 \right|}

\newcommand{\be}{\begin{equation}}
\newcommand{\ee}{\end{equation}}
\newcommand{\bea}{\begin{eqnarray}}
\newcommand{\eea}{\end{eqnarray}}

\usepackage{dcolumn}
\usepackage{bm}
\usepackage{amssymb,amsmath}
\usepackage{color}
\usepackage{float}

\definecolor{DarkGreen}{rgb}{0,0.6,0.2}

\begin{document}
\title{Two-Photon Interference and Entanglement in Coupled Cavities}
\author{Imran M. Mirza and S.J. van Enk}
\affiliation{Oregon Center for Optics and Department of Physics\\University of Oregon\\
Eugene, OR 97401}
\begin{abstract}
Two-photon interference effects, such as the Hong-Ou-Mandel (HOM) effect, can be used to characterize to what extent two photons are identical. Identical photons are necessary for both linear optics quantum computing and single-photon quantum cryptography. We study here how storage and delay of photons in coupled cavity arrays, which inevitably will lead to changes in the photons' spectral and temporal profiles, affects their HOM interference. In addition we consider various types of entanglement that occur naturally in such a context.
\end{abstract}
  \maketitle
\section{Introduction}
The Hong-Ou-Mandel (HOM) effect \cite{HOM} is a celebrated example of a pure {\em quantum} interference effect: the fact that two photons impinging on  the two input ports of a 50/50 beam splitter always emerge together in one output port is {\em not} affected by any phase shifts (where a phase shift is to be  distinguished from a time delay, see \cite{van2011phase}: a time delay does affect both Hong-Ou-Mandel and classical interference.) 

applied prior to impinging, unlike in the case of interference of two classical fields on the same 50/50 beam splitter.
The destructive interference between the two paths that lead to the same final state with both photons exiting in different output ports
can be perfect only if at the output the two photons are indistinguishable. They must, in particular, have identical spectral and polarization states at the output.
In principle there is no such requirement for the photons at the input, and HOM-like interference can occur, for example, between photons of different colors as well \cite{RaymerHOM}, provided there is a frequency-changing mechanism between input and output.

In the present paper we consider two-photon interference effects in the context of coupled cavity arrays. There has been great interest in such arrays in the last dozen years or so because of their ability to delay and store light \cite{yariv1999coupled,heebner2002scissor,heebner2002slow,scheuer2005coupled,baba2008slow,krauss2008we}.
Most research has focused on classical light but 
storing single photons is important for quantum communication purposes, too, in particular for
entanglement purification and quantum repeaters, which promise to increase the distance over which quantum key distribution can be securely employed \cite{qrepeat,qkd}.
For linear optics quantum computing as well as for quantum cryptographic purposes it is crucial that spectral and temporal lineshapes of photons are not distorted by the storing and retrieval process.
The two-photon interference effects we study here provide a sensitive test for such unwanted distortion effects \cite{legero2006, specht2009phase}. In addition, entanglement is sensitive to coherence properties of the photons, and we investigate that aspect here, too.

Just as we did in our recent work \cite{Imran} on single-photon effects in coupled cavity arrays, we will include the generation of the two photons explicitly, by assuming we have two single emitters (which could be single atoms or single quantum dots or NV centers in diamond \cite{santori2002indistinguishable,mckeever2004deterministic,englund2010,riedrich2011}), one in each of two cavities. Unlike that work (and almost all of related work on coupled cavities) we do not assume a uni-directional coupling. Instead, the two photons can travel back and forth between the two cavities.
This symmetry between the two photons and the two propagation directions leads one to expect HOM-like interference effects. Since many more processes are occurring in our setup than do in the standard HOM setup, especially nonlinear optics effects due to the presence of the two atoms, one would expect in our case the interference effects to be less pronounced
and more complicated \cite{nazir2009overcoming}.

As mentioned above, we will discuss entanglement as well. Because our system consists of four cavity modes and two atoms distributed evenly over
two locations, bipartite entanglement of different types can occur: between the cavity modes, between the atoms, and hybrid entanglement between atom and cavity modes.
For example, mode entanglement may be of the  form  $(\ket{0}_{a_{1}}\ket{1}_{a_{2}}\ket{1}_{a_{3}}\ket{0}_{a_{4}}
-\ket{1}_{a_{1}}\ket{0}_{a_{2}}
\ket{0}_{a_{3}}\ket{1}_{a_{4}})/\sqrt{2}$, which can be interpreted as entanglement between the two photons (one on each side; the subscripts here indicate the four counter propagating modes in the two cavities as indicated in FIG.~1), or, alternatively, of the form
$(\ket{0}_{L}\ket{2}_{R}-\ket{2}_{L}\ket{0}_{R})/\sqrt{2}$, which occurs in the HOM effect \cite{Megan}, and which cannot be interpreted as entanglement between the two photons (instead, it is the modes that are entangled \cite{enk03}). In this state subscripts are showing the left (L) and right (R) atom-cavity systems as shown in FIG.~1.

The paper is organized as follows.
We describe our system and the theoretical methods we employ in Section II. The description of unidirectional coupling can be done elegantly within the formalism of
quantum cascaded systems combined with quantum trajectories. In our case we can still straightforwardly use the latter, but the former theory has to be adjusted to account for bidirectional coupling. With the help of these methods, we study two-photon interference effects in Section III, and in Section IV we analyze the various types of entanglement occurring in our system.

\section{Two spatially separated atom-cavity systems}
\subsection{Model and Hamiltonian}
\begin{figure*}[t]
\includegraphics[width=6in,height=2.4in]{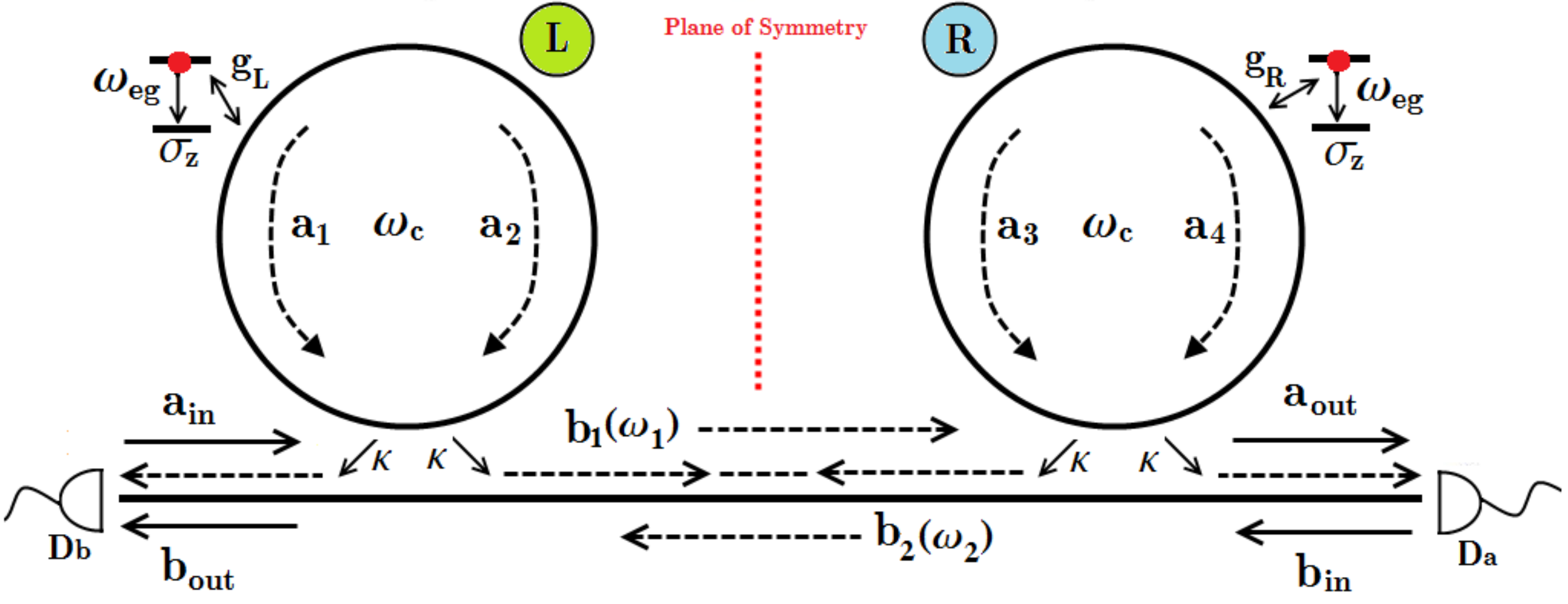}
\captionsetup{
  format=plain,
  margin=1em,
  justification=raggedright,
  singlelinecheck=false
}
  \caption{Two spatially separated atom-cavity systems, and two single-photon detectors. Thanks to the bi-directional coupling between the two cavities, excitations can be transfered between the atom-cavity systems multiple times before being detected. We consider here a mirror-symmetric system, with all coupling constants, decay rates, and resonance frequencies pairwise the same for the left and right atom-cavity systems.
The detectors count photons in the two output modes, described by annihilation operators $\hat{a}_{out}$ and $\hat{b}_{out}$. For further details, see main text.
    }\label{Fig1}
 \end{figure*}
We have two spatially separated atom-cavity systems (referred to as ``left'' or ``L'' and ``right'' or ``R'', respectively) coupled through an optical fiber which is assumed to have two continua of modes (propagating to the left and right, respectively), as shown in FIG.~1. A single photon is generated in each cavity through an initially excited atom (with transition frequency $\omega_{eg}$: both atoms are taken to be identical in the rest of the paper and the spontaneous emission from the atoms is set to zero). Due to the atom-cavity coupling (represented by complex coupling coefficients $g_{L}$ and $g_{R}$ for left and right systems, respectively) the emitted photon can excite any one of the two counter propagating cavity modes, which are described by annihilation operators $\hat{a}_{1}$, $\hat{a}_{2}$ for the left cavity and $\hat{a}_{3}$, $\hat{a}_{4}$ for the right cavity. Inside each cavity, both modes are assumed to have the same single resonant frequency $\omega_{c}$.

There are two possibilities for the excitation to leak out of a given cavity. For example, for the left cavity, the photon in the mode $\hat{a}_{2}$ can exit towards the left (at a leakage rate $\kappa$) and will be detected by detector $D_{b}$.
 On the other hand, if the photon is in the mode $\hat{a}_{1}$, then it can escape towards the right (at the same leakage rate $\kappa$), after which it can enter into the right cavity due to the evanescent coupling between fiber and  cavity. It may, alternatively, go straight to the detector $D_a$.
Excitations can shuttle back and forth many times before finally being lost by the system and detected by the two detectors.
 
In our system there is a time delay $\tau$ between the cavities (which is defined in terms of the separation $d$ between cavities as $\tau =d/c$, with $c$ the group velocity of light in the fiber, which is assumed to be constant around the cavities' and atoms' resonant frequencies). Such time delays appear in the context of cascaded quantum networks \cite{carmichael1993quantum,gardiner1993driving} where they are considered arbitrary constants that can be eliminated, since they prove irrelevant to the physics of the problem. But for our system  we cannot so simply ignore the time delay. This is due to the fact that the coupling between system L and R is not unidirectional. From this perspective our model resembles more a quantum feedback network \cite{wiseman2009quantum,gough2009quantum}, with the difference that there is no special part added to the actual system to perform this feedback \cite{petersen2011cascade,vitali2000quantum}. Rather, this happens due to the geometry of the system itself.

Assuming no coupling between the intra cavity modes and applying the standard rotating wave (RWA) and Markov approximations, the Hamiltonian of the global system (atoms, cavities and the fiber) takes the following form:
 \begin{widetext}
\begin{equation}\label{H}
\begin{split}
& \hat{H}=-\hbar\omega_{eg}\hat{\sigma}^{(L)}_{-}\hat{\sigma}^{(L)}_{+}-\hbar\omega_{eg}\hat{\sigma}^{(R)}_{-}\hat{\sigma}^{(R)}_{+}+ 
\hbar\omega_{c}(\hat{a}_{1}^{\dagger}\hat{a}_{1}+ \hat{a}_{2}^{\dagger}\hat{a}_{2}
+\hat{a}_{3}^{\dagger}\hat{a}_{3}+ \hat{a}_{4}^{\dagger}\hat{a}_{4})
+\hbar(g_{L}\hat{a}_{1}^{\dagger}\hat{\sigma}^{(L)}_{-}+g^{\ast}_{L}\hat{a}_{1}\hat{\sigma}^{(L)}_{+}) +\hbar(g^{\ast}_{L}\hat{a}_{2}^{\dagger}\hat{\sigma}^{(L)}_{-}+g_{L}\hat{a}_{2}\hat{\sigma}^{(L)}_{+})\\
&\hspace{7.5mm}+\hbar(g_{R}\hat{a}_{3}^{\dagger}\hat{\sigma}^{(R)}_{-}+g^{\ast}_{R}\hat{a}_{3}\hat{\sigma}^{(R)}_{+}) +\hbar(g^{\ast}_{R}\hat{a}_{4}^{\dagger}\hat{\sigma}^{(R)}_{-} +g_{R}\hat{a}_{4}\hat{\sigma}^{(R)}_{+})
+\hbar\int_{-\infty}^{+\infty}\omega_{1} \hat{b}_{1}^\dagger(\omega_{1})\hat{b}_{1}(\omega_{1})d\omega_{1}
+\hbar\int_{-\infty}^{+\infty}\omega_{2} \hat{b}_{2}^\dagger(\omega_{2})\hat{b}_{2}(\omega_{2})d\omega_{2}\\
&\hspace{7mm}+i\hbar\sqrt{\frac{\kappa}{2\pi}}\int_{-\infty}^{+\infty}\Bigg(\hat{a}_{1}\hat{b}_{1}^{\dagger}(\omega_{1})-\hat{a}_{1}^{\dagger}\hat{b}_{1}(\omega_{1})+\hat{a}_{3}\hat{b}^{\dagger}_{1}(\omega_{1})
-\hat{a}_{3}^{\dagger}\hat{b}_{1}(\omega_{1})\Bigg)d\omega_{1}
+i\hbar\sqrt{\frac{\kappa}{2\pi}}\int_{-\infty}^{+\infty}\Bigg(\hat{a}_{2}\hat{b}_{2}^{\dagger}(\omega_{2})-\hat{a}_{2}^{\dagger}\hat{b}_{2}(\omega_{2})\\
&\hspace{7.5mm}+\hat{a}_{4}\hat{b}^{\dagger}_{2}(\omega_{2})-\hat{a}_{4}^{\dagger}\hat{b}_{2}(\omega_{2})\Bigg)d\omega_{2}.
\end{split}
\end{equation}
\end{widetext}
Here $\hat{\sigma}^{(L)}_{+},\hat{\sigma}^{(R)}_{+}$ are the atomic raising operators for left and right atoms respectively and $\hat{b}_{1}(\omega_{1})$, $\hat{b}_{2}(\omega_{2})$ are the annihilation operators for two fiber continua. The nonvanishing commutation relations are: $[\hat{\sigma}^{(L)}_{+},\hat{\sigma}^{(L)}_{-}]=\hat{\sigma}^{(L)}_{z}$ and a similar relation for right atom, $[\hat {b}_{i}(\omega_{i}),\hat {b}_{j}^{\dagger}(\omega_{j}))]=\delta(\omega_{i}-\omega_{j})$ $\forall i=1,2$; $j=1,2$ and $[\hat {a}_{i},\hat {a}_{j}^{\dagger}]=\delta_{ij}$ $\forall i=1,2,3,4$; $j=1,2,3,4$. We have chosen the energy of the atomic ground states to be negative (first two terms), such that the initial state has zero energy.

The interaction of the intra cavity modes with the fiber continua makes both left and right systems open and to describe the dynamics of such an open system we now transform to the Heisenberg picture. Following the standard procedure \cite{gardiner2004quantum, carmichael2008statistical} of eliminating continua in the Heisenberg picture and identifying the two input operators corresponding to two continua:
\begin{equation}
\begin{split}
&\hat{a}_{in}(t)=\frac{1}{\sqrt{2\pi}}\int_{-\infty}^{\infty}\hat{b}_{1}(\omega_{1})e^{i\omega_{1}(t-t_{0})}d\omega_{1}\\
&\hat{b}_{in}(t)=\frac{1}{\sqrt{2\pi}}\int_{-\infty}^{\infty}\hat{b}_{2}(\omega_{2})e^{i\omega_{2}(t-t_{0})}d\omega_{2}
\end{split}
\end{equation}
we finally arrive at the following Quantum Langevin's equation for an arbitrary system operator $\hat{X}(t)$ (which can either belong to system L or to system R):
\begin{widetext}
\begin{equation}\label{Lang}
\begin{split}
& \frac{d\hat{X}(t)}{dt}=-\frac{i}{\hbar}[\hat{X}(t),\hat{H}_{s}]\\
& \hspace{13mm}-[\hat{X}(t),\hat{a}_{1}^\dagger]\Bigg(\frac{\kappa}{2}\hat{a}_{1}+\sqrt{\kappa}\hat{a}_{in}(t)\Bigg)+\Bigg(\frac{\kappa}{2}\hat{a}_{1}^{\dagger}+\sqrt{\kappa}\hat{a}_{in}^{\dagger}(t)\Bigg)[\hat{X}(t),\hat{a}_{1}]
-[\hat{X}(t),\hat{a}_{3}^\dagger]\Bigg(\frac{\kappa}{2}\hat{a}_{3}+\sqrt{\kappa}\hat{a}_{in}(t-\tau)\Bigg)\\
&\hspace{13mm}+\Bigg(\frac{\kappa}{2}\hat{a}_{3}^{\dagger}+\sqrt{\kappa}\hat{a}_{in}^{\dagger}(t-\tau)\Bigg)[\hat{X}(t),\hat{a}_{3}]
-\kappa[\hat{X}(t),\hat{a}_{3}^{\dagger}]\hat{a}_{1}(t-\tau)+\kappa\hat{a}_{1}^{\dagger}(t-\tau)[\hat{X}(t),\hat{a}_{3}]\\
&\hspace{13mm}-[\hat{X}(t),\hat{a}_{2}^\dagger]\Bigg(\frac{\kappa}{2}\hat{a}_{2}+\sqrt{\kappa}\hat{b}_{in}(t-\tau)\Bigg)+
\Bigg(\frac{\kappa}{2}\hat{a}_{2}^{\dagger}+
\sqrt{\kappa}\hat{b}^{\dagger}_{in}(t-\tau)\Bigg)[\hat{X}(t),\hat{a}_{2}]
-[\hat{X}(t),\hat{a}_{4}^\dagger]\Bigg(\frac{\kappa}{2}\hat{a}_{4}+ \sqrt{\kappa}\hat{b}_{in}(t)\Bigg)\\
&\hspace{13mm}+\Bigg(\frac{\kappa}{2}\hat{a}_{4}^{\dagger}+\sqrt{\kappa}\hat{b}_{in}^{\dagger}(t)\Bigg)[\hat{X}(t),\hat{a}_{4}]
-\kappa[\hat{X}(t),\hat{a}_{2}^{\dagger}]\hat{a}_{4}(t-\tau)
+\kappa\hat{a}_{4}^{\dagger}(t-\tau)[\hat{X}(t),\hat{a}_{2}].
\end{split}
\end{equation}
\end{widetext}
Here $\hat{H}_{s}$ is the atom-cavity system Hamiltonian, which consists of the discrete terms in Eq.~[\ref{H}]. The above Langevin equation is a generalization of the usual cascaded quantum system Langevin equation \cite{gardiner1993driving,gardiner2004quantum} to include a bidirectional coupling between left and right systems. Corresponding to two input field operators $\hat{a}_{in}$, $\hat{b}_{in}$  appearing in the above equation there are two output operators $\hat{a}_{out}$, $\hat{b}_{out}$ which are related to the input operators and the intra cavity field operators through the input-output relations \cite{carmichael2008statistical,gardiner1985,carmichael1993open} as 
\begin{subequations}
\begin{eqnarray}
\hat{a}_{in}^{(R)}(t)=\hat{a}_{{\rm out}}^{(L)}(t-\tau)=\hat{a}_{{\rm in}}^{(L)}(t-\tau)+\sqrt{\kappa}\hat{a}_{1}(t-\tau),\\
\hat{b}_{in}^{(L)}(t)=\hat{b}_{{\rm out}}^{(R)}(t-\tau)=\hat{b}_{{\rm in}}^{(R)}(t-\tau)+\sqrt{\kappa}\hat{a}_{4}(t-\tau).
\end{eqnarray}
\end{subequations}
Note that the output from one cavity is serving as the input to the other cavity (with the delay time included), so that the coupling is explicitly bidirectional. We have also explicitly included (redundant) L and R superscripts here to make the distinction among the various input and output operators more transparent. The nonvanishing commutation relations among the input operators are given by:
$[\hat{a}_{{\rm in}}(t),\hat{a}_{{\rm in}}^{\dagger}(t')]=\delta(t-t')$, $[\hat{b}_{{\rm in}}(t),\hat{b}_{{\rm in}}^{\dagger}(t')]=\delta(t-t')$. 

If we denote by $\ket{\Psi}$ the initial state of the global system (atoms, cavities and fiber), we have $\hat{a}_{{\rm in}}\ket{\Psi}=0$ and $\hat{b}_{{\rm in}}\ket{\Psi}=0$, as initially there is no photon present. These input operators, therefore, do not contribute to the expectation values of normally ordered observables.

Although the time delay arising from the fiber cannot be ignored due to the feedback mechanism in our system, for the present study we are more interested in the delays caused by the excitations remaining inside the cavities. (In fact, the whole point of using coupled cavity arrays is to store and delay photons inside cavities.) This cavity-induced time delay is on the order of $\kappa^{-1}$ and under the condition that $\kappa\tau<<1 $ we can in fact ignore the trivial delay $\tau$. From now on we are going to focus on this particular regime---the experimentally relevant regime---and we set $\tau\rightarrow 0$ for that reason.

\subsection{Quantum trajectory analysis}
Now we transform back to the Schr\"odinger picture and make use of the Quantum Trajectory Method (or quantum jump method) \cite{carmichael1993open,dum1992monte, molmer1993monte} which is an appropriate formalism for the description of open quantum systems. This analysis applied to the system under study implies that during any (infinitesimally) small time interval we have one of two possibilities: either a photon leaks out of the system and one of the detectors registers it (and so a quantum jump takes place), or the excitation(s) remain inside the system and no jump is recorded. The next subsections are devoted to the detailed study of both these situations.

\subsubsection{Occurrence of a jump}
In the Quantum Trajectory Method, photodetection at the output ports is described by the output operators (also called jump operators in this context), which in our case are denoted by $\hat{J}_{a}=\hat{a}_{{\rm out}}$ and $\hat{J}_{b}=\hat{b}_{{\rm out}}$. Detector $D_{a}$ detects the field $\hat{a}_{{\rm out}}$ and $D_{b}$ detects the field $\hat{b}_{{\rm out}}$ (see Fig.[\ref{Fig1}]). The detection events happen at random times with certain probabilities determined by the jump/output operators $\hat{J}_j$ for $j=a,b$, and by the current state $\ket{\psi}$. During an infinitesimal time interval $[t,t+dt]$ the detection probability is given by
\begin{equation}\label{pi}
P_{j}(t)={\bra{\psi}}\hat{J_{j}}^{\dagger}
\hat{J_{j}}{\ket{\psi}}dt=:\Pi_j dt,
\end{equation}
for $j=a,b$. After one jump is recorded we have to reset the state according to the transformation:
\begin{equation}
\ket{\psi}\mapsto
\frac{\hat{J}_j\ket{\psi}}{\sqrt{\Pi_j}}.
\end{equation}
The normalization factor $\Pi_j$ appearing here is in fact the probability density defined in Eq.~(\ref{pi}).

\subsubsection{Non-unitary evolution}
According to the Quantum Trajectory Method, when no detector clicks, the system dynamics follows a non-unitary evolution described by a non-unitary Schr\"odinger equation:
\begin{equation}\label{NUSE}
i\hbar\frac{d\ket{\tilde{\psi}(t)}}{dt}=\hat{H}_{NH}{\ket{\tilde{\psi}(t)}}.
\end{equation}
The ``Non-Hermitian Hamiltonian'' $\hat{H}_{NH}$ appearing in the above equation turns out to be the sum of the standard (Hermitian) system Hamiltonian (Eq.~[\ref{H}]) and an anti-Hermitian term constructed from the jump operators, such that
\begin{equation}\label{NHH}
\hat{H}_{NH}=\hat{H}_{s}-i\sum_{j=a,b}\hat{J}_j^\dagger\hat{J}_j/2.
\end{equation}
The unnormalized ket $\ket{\tilde{\psi}(t)}$ is called the ``No-Jump state," which is a pure state whose norm decays in time. It can be written as a linear combination of all the different possibilities of finding all excitations in the system that have not been detected yet.

\section{Two-photon quantum interference effects}
In this Section we will analyze two-photon interference effects. In particular, we study whether the probability to detect the two photons in the same detector differs from the probability to detect them in different detectors.
We consider two cases: first a case of mere theoretical significance where we compare joint detection probabilities in a small time interval (so the photons are detected at the same time), and second a case of experimental relevance where one records at what detectors and at what times the two photons were detected.

Some aspects of  single photon transmission could be derived using a semi-classical approach (see for instance \cite{srinivasan2007mode}, \cite{elyutin2012interaction}).
Here, on the other hand, we are interested in interference of the Hong-Ou-Mandel type, which cannot be explained semi-classically \cite{mandel}, and we thus
follow the procedure outlined in the preceding Section. The ``No-Jump state'' $\ket{\tilde{\psi}(t)}$ describing the situation where neither excitation has been detected yet,  consists of a superposition of 19 different states, corresponding to the 19 different ways of finding the two excitations in the different parts of the system. We write 
\begin{equation}\label{nojumptotal}
\begin{split}
&\ket{\tilde{\psi}(t)}=c_{1}(t)\ket{e_{1}00,e_{2}00}+c_{2}(t)\ket{e_{1}10,g_{2}00}\\
&+c_{3}(t)\ket{e_{1}01,g_{2}00}+c_{4}(t)\ket{e_{1}00,g_{2}10}+c_{5}(t)\ket{e_{1}00,g_{2}01}\\
&+c_{6}(t)\ket{g_{1}10,e_{2}00}+c_{7}(t)\ket{g_{1}01,e_{2}00}+c_{8}(t)\ket{g_{1}00,e_{2}10}\\
&+c_{9}(t)\ket{g_{1}00,e_{2}01}+c_{10}(t)\ket{g_{1}20,g_{2}00}+c_{11}(t)\ket{g_{1}02,g_{2}00}\\
&+c_{12}(t)\ket{g_{1}00,g_{2}20}+c_{13}(t)\ket{g_{1}00,g_{2}02}+c_{14}(t)\ket{g_{1}11,g_{2}00}\\
&+c_{15}(t)\ket{g_{1}10,g_{2}10}+c_{16}(t)\ket{g_{1}10,g_{2}01}+c_{17}(t)\ket{g_{1}01,g_{2}10}\\
&+c_{18}(t)\ket{g_{1}01,g_{2}01}+c_{19}(t)\ket{g_{1}00,g_{2}11}.
\end{split}
\end{equation}
The notation we used here is as follows: the first slot in the ket is the state of the left atom and the next two slots display the number of photons in the modes of the left cavity. The remaining three slots are for the right system with the atomic and cavity states ordered in the same way. 
\subsection{Photons detected at the same time}
We study the interference effects in our system by calculating the joint probabilities of detecting the two photons at the output ports. Here we remind the reader that we are working in the regime where trivial fiber delays are neglected, and so one type of interference (of a theoretical nature) can be studied by considering the equal-time probability densities. We thus compare 
\begin{equation}
\begin{split}\label{P2}
&\text{Probability density of getting two clicks at the same }\\
& \text{time t at detector $D_{a}$}\equiv P_{2}= \langle \tilde{\psi}(t) \vert \hat{a}_{out}^{\dagger 2}\hat{a}_{out}^{2} \vert \tilde{\psi}(t) \rangle {\rm \delta T}\\
&\hspace{20mm} =\kappa^{2}\Bigg\vert \sqrt{2}c_{10}(t)+\sqrt{2}c_{12}(t)+2c_{15}(t)\Bigg\vert^{2}{\rm \delta T}
\end{split}
\end{equation}
(with $\delta T$ is a very small time interval compared to the cavity leakage time $\kappa^{-1}$) 
\begin{figure*}
\begin{center}
\begin{tabular}{cccc}
\subfloat{\includegraphics[width=9cm,height=7cm]{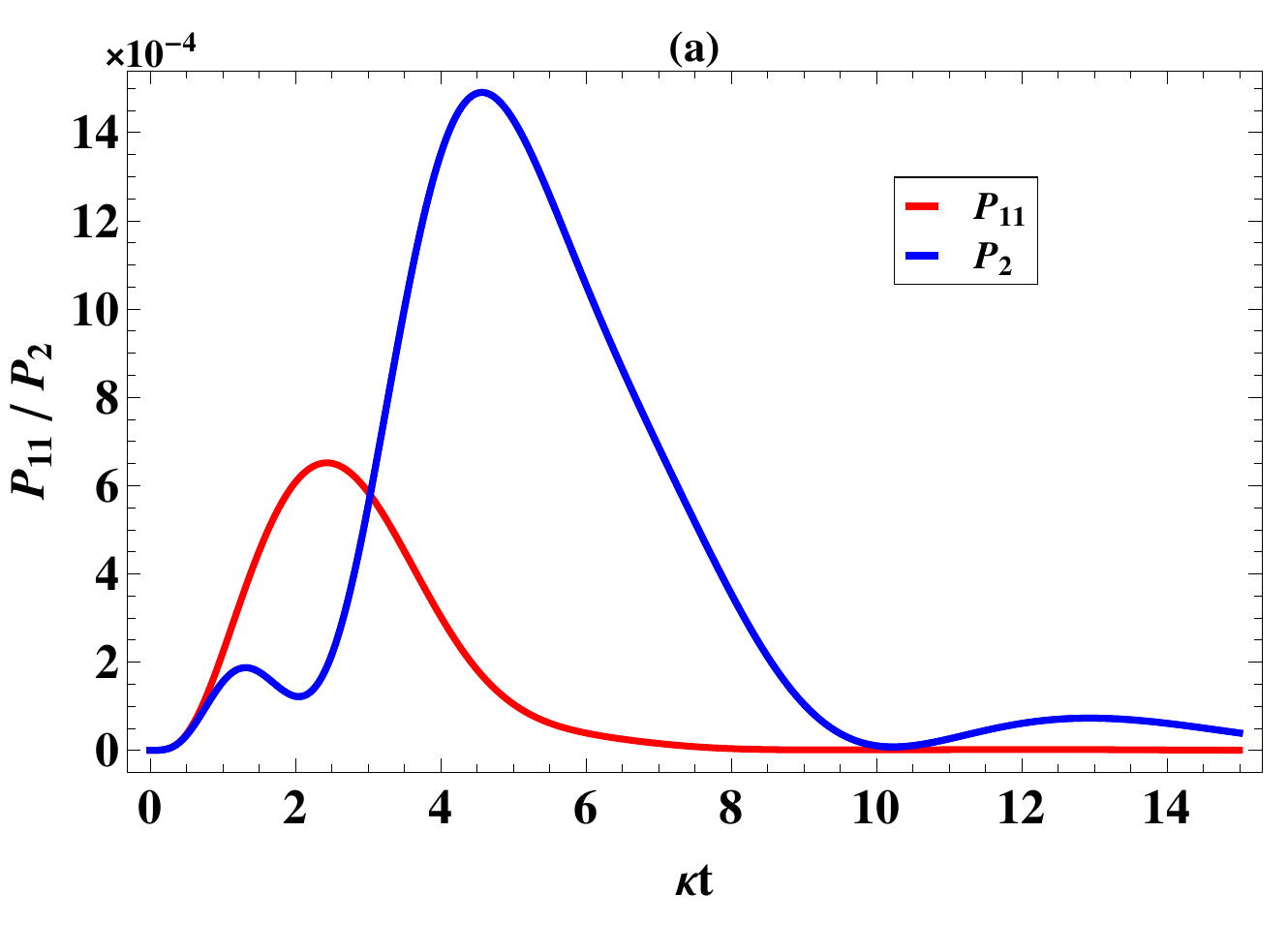}} & 
\subfloat{\includegraphics[width=9cm,height=7cm]{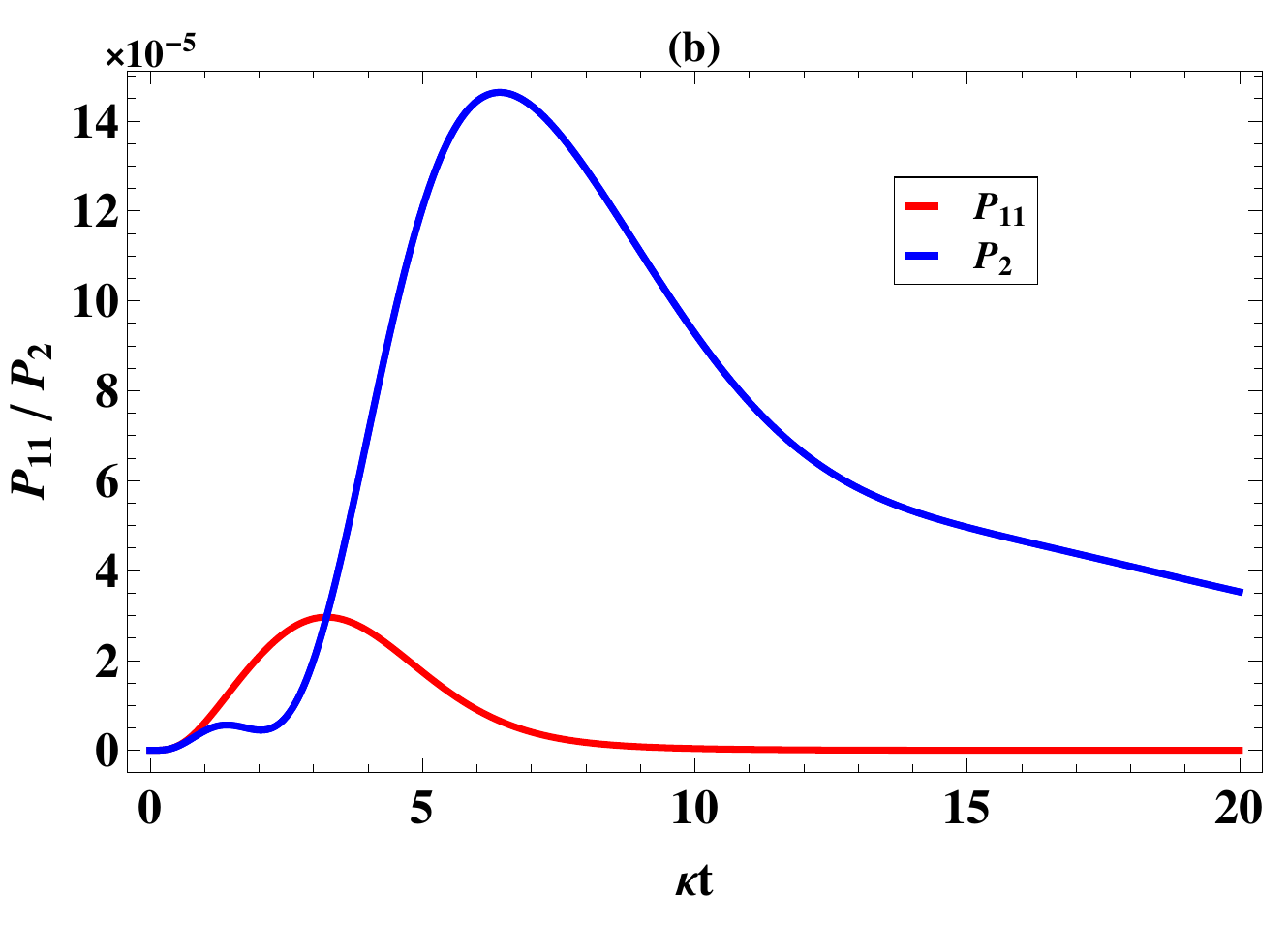}}\\
\end{tabular}
\captionsetup{
  format=plain,
  margin=1em,
  justification=raggedright,
  singlelinecheck=false
}
\caption{Joint probability densities of detecting photons at the output ports as functions of time. We assume a weak coupling regime with $|g_{L}|=|g_{R}|\equiv|g|$. In
(a) we choose $|g|/\kappa = 0.25, \Delta/\kappa=0.5, {\rm \delta T = 0.1\kappa^{-1} }$, and in 
(b) we choose  $|g|/\kappa=0.1$ with all other parameters the same as in part (a). It is more likely one detects photons at the same output, than at different outputs, reminiscent of the Hong-Ou-Mandel effect.}\label{Fig2}
\end{center}
\end{figure*}
with
\begin{equation}\label{P11}
\begin{split}
&\text{Probability density of getting one click at detector $D_{a}$}\\
& \text{and the other at detector $D_{b}$ at the same time t}\\
& \hspace{5mm}\equiv P_{11}= \langle \tilde{\psi}(t) \vert \hat{b}_{out}^{\dagger}\hat{a}_{out}^{\dagger}\hat{a}_{out}\hat{b}_{out} \vert \tilde{\psi}(t) \rangle {\rm \delta T}\\
& \hspace{15mm}=\kappa^{2}\Bigg\vert c_{14}(t)+c_{16}(t)+c_{17}(t)+c_{19}(t)\Bigg\vert^{2}{\rm \delta T}
\end{split}
\end{equation}
We will call the latter the ab/ba detection density and the former the aa/bb detection density (here we use the fact that because of the mirror symmetry we imposed, the joint probability of getting two clicks at detector a is the same as that of getting two clicks at detector b). 

We plot both densities in FIG.~2 in the weak coupling regime ($(|g_{L}|,|g_{R}|) < \kappa$). The reason for us focusing on the weak coupling regime is that in this regime photons are more likely to leak out from the cavities rather than being reabsorbed by the atoms, and hence the chances of observing equal-time interference are larger than in the strong coupling regime, in which excitations can go through several rounds of absorption and emission by the atoms, before finally being detected.

In part (a) of the figure we have chosen $|g_{L}|=|g_{R}|\equiv|g|=\kappa/4$,  and all coupling rates are real and positive. The curve describing the ab/ba detection (red curve) shows a single maximum. For initial times this primarily indicates the two processes where the single photons from the L and R systems escape directly towards the detectors, either on their own side, or on the other side, immediately after the de-excitation of the atoms in their respective cavities. All other processes take a longer time to deliver both photons at the detectors.

In comparison to the red curve, the blue curve representing aa/bb detection at the same detector shows two maxima due to interference of the different ways the same final situation can occur. For example, one possibility is that one photon in the L cavity mode $\hat{a}_{1}$ mode escapes the cavity and is directly detected by $D_{a}$ (never entering the R cavity) while the other photon joins it after having escaped the R cavity through the $\hat{a}_{3}$ mode. Since the fiber delay can be neglected both these photons will be detected at about the same time. But another possibility leading to both photons being detected at detector a is that the photon emitted by the left cavity actually enters the other cavity first, is reabsorbed by the atom, and is then reemitted in the reverse direction. Clearly, such a process takes a longer time, on the order of 2 cavity decay times plus $g^{-1}=4\kappa^{-1}$, which equals about 6 $\kappa^{-1}$.
The  quantum interference of all such possible routes (some taking longer than others) to the same final state generates the blue curve.
 
In FIG.2~(b) we have chosen the value of $|g|$ to be one tenth of cavity decay rate, thus going to an even weaker coupling scenario. The main effect of this change is that now the  destructive interference of the HOM type is stronger in this case (the red curve is lower, the blue curve is higher), as the the nonlinear processes involving the atoms are less likely to occur (for the actual HOM effect there are no nonlinear processes at all).

\begin{figure}[h]
\begin{center}
\includegraphics[width=9cm,height=6.5cm]{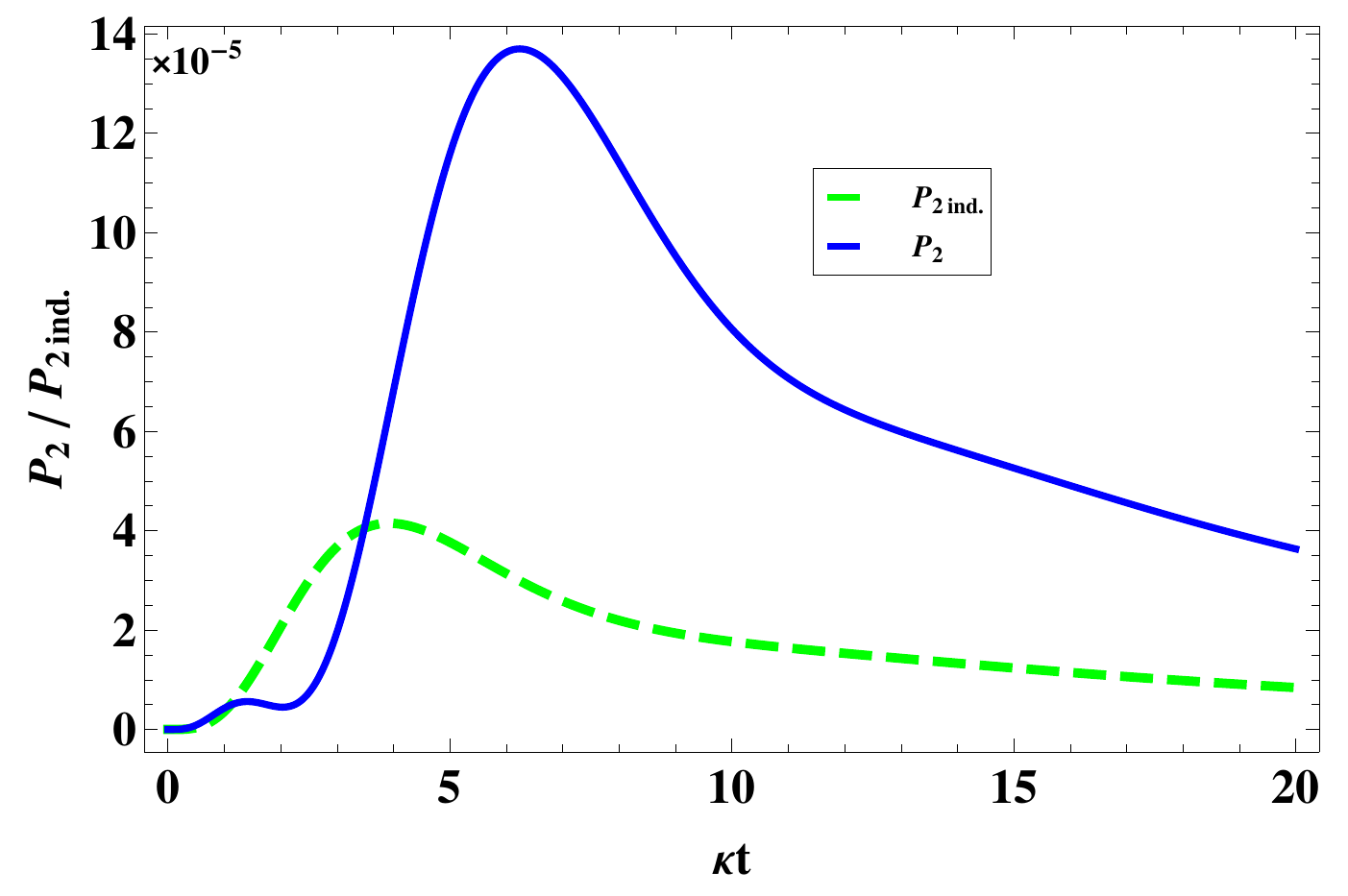}
\captionsetup{
  format=plain,
  margin=1em,
  justification=raggedright,
  singlelinecheck=false
}
\caption{Equal-time probability density for detecting photons at the output  for two  independent atom-cavity systems, in the weak coupling regime (dashed green curve). We chose $ |g|/\kappa=0.1, \Delta/\kappa=0.5, {\rm \delta T = 0.1\kappa^{-1} }$. For comparison we replotted the blue curve from FIG.~2(b).}\label{Fig3}
\end{center}
\end{figure}

In order to separate out interference effects, we now consider, by way of comparison,
a fictitious system consisting of two independent cavities that cannot display any interference: the probability of registering two clicks at two given times is simply the product of the probabilities of one system to emit a photon at those times. In FIG.~\ref{Fig3} we plot the equal-time probability density, and compare it to the aa/bb probability density. We notice that the detection probability for the fictitious case shows a single maximum and no oscillations (compare this with FIG.~2). This is an indication that interference effects are missing in the independent cavity case. Also note that the graph  resembles the case of two coupled cavities and detections at different detectors, confirming that in the latter case chances of interference are small.

\subsection{Photons detected at arbitrary times}
The probability of getting two clicks at more or less the same time (as defined in Eq.~[\ref{P2}] and Eq.~[\ref{P11}] respectively) will be very small. A more likely event is that we detect the photons at different times.
We address this situation by considering a simulation of a feasible experiment. The experiment records at what times which detectors click, and the analysis of the data is then supposed to reveal the presence of quantum interference. The latter ought to be manifested in differences between the distributions of waiting times between clicks at the same detector and waiting times between clicks at different detectors.

We performed a Quantum Monte Carlo simulation consisting of over 20,000 trajectories, and we recorded the times at which the two photons are detected at the outputs. We use the following convention: time $T_{1}$ indicates the time of arrival of the first detected click, and $T_{2}$ that of the second.  By this definition $T_{2}>T_{1}$. 

It turns out that for the parameters used in FIG.~\ref{Fig2}, 62 percent of the trajectories lead to clicks at the same detector (and those events end up in plot (a)), while in 38\% of the cases photons are detected at different detectors (plot (b)).
This imbalance is a clear indication of HOM-type interference. One can also discern that the waiting time between first and second clicks tends to be larger for the ab/ba case than for the aa/bb case. So, in the latter case the photons tend to bunch together, just as in the HOM effect.

\begin{figure}
\begin{center}
\begin{tabular}{cccc}
\subfloat{\includegraphics[width=4.3cm,height=8cm]{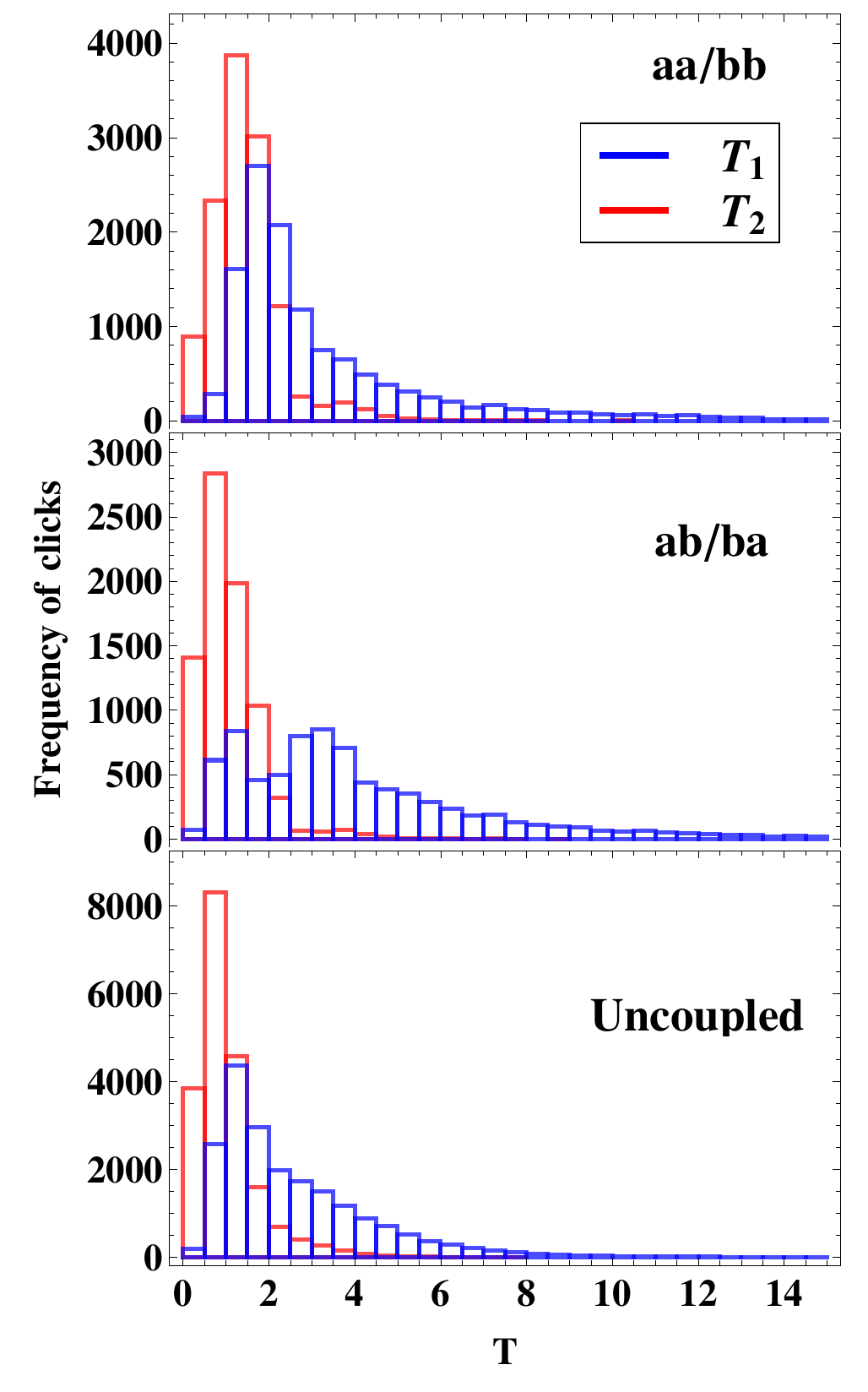}} & 
\subfloat{\includegraphics[width=4.3cm,height=8cm]{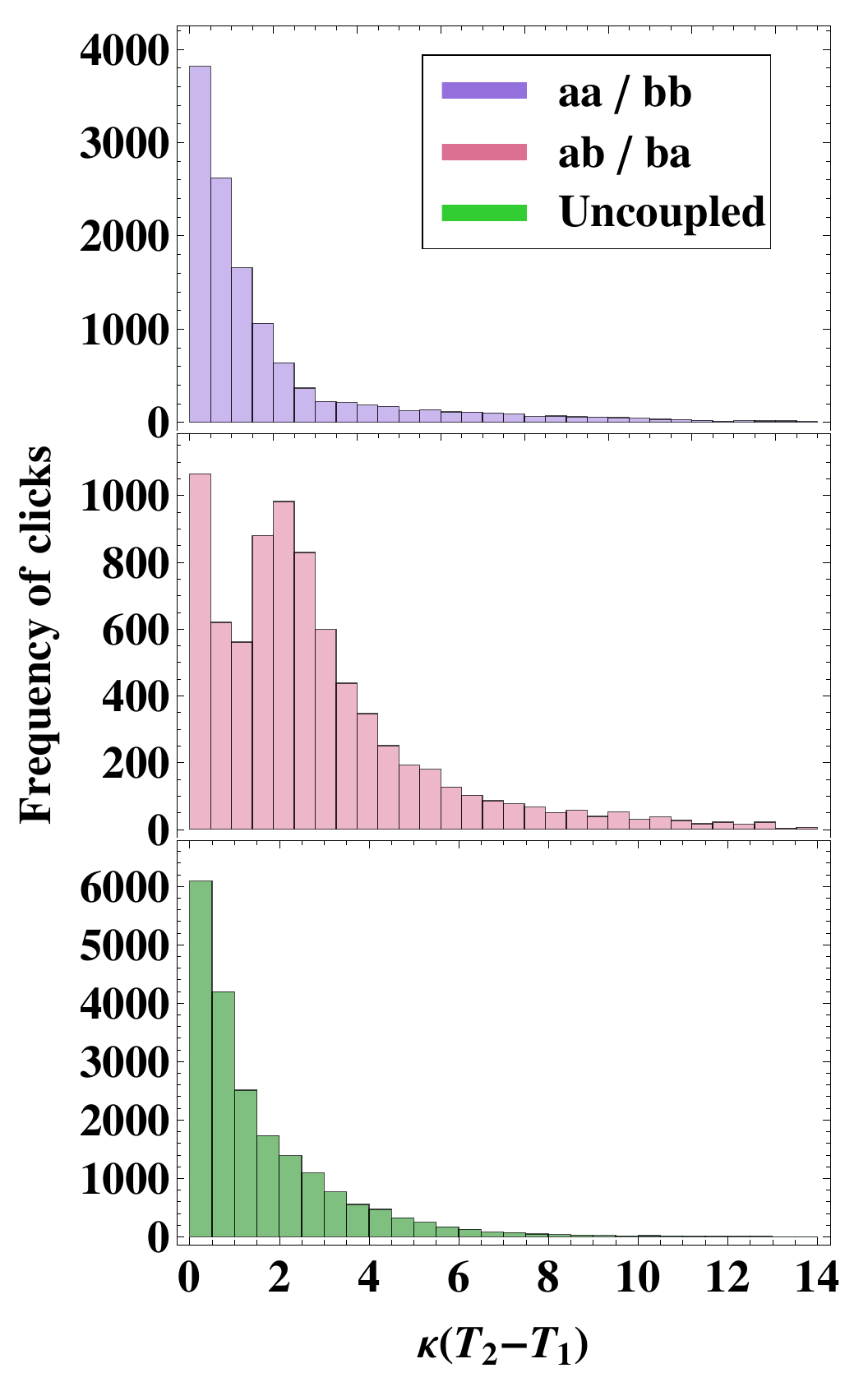}}\\
\end{tabular}
\captionsetup{
  format=plain,
  margin=1em,
  justification=raggedright,
  singlelinecheck=false
}
\caption{Frequency histograms of detection times (left column) and detection time differences (right column). The widths of the bins of the histograms are chosen to be $0.5\kappa$. The top figures in each column refer to $aa/bb(T_{1},T_{2})$, the middle figures correspond to $ab/ba(T_{1},T_{2})$ and the bottom figures to two independent atom-cavity systems.}\label{Fig4}
\end{center}
\end{figure}

In FIG.~\ref{Fig4} we have plotted the histograms of the individual detection times, as well as the time differences $T_2-T_1$. From the plots it is clear that for $aa/bb(T_{1},T_{2})$ detections that time difference is typically shorter than that for the $ab/ba(T_{1},T_{2})$ detections, thus confirming the observation we made concerning the previous Figure. We also note the presence of destructive interference in the $ab/ba(T_{1},T_{2})$ detection case around a time $2\kappa^{-1}$. 

\subsection{Effects of angular position on interference}
We have chosen to study a mirror-symmetric system, so as to maximize the possibility of HOM-type interference between two photons.
That mirror-symmetry in our system implies that if the left atom-cavity coupling rate is  $g$ then the right atom-cavity  coupling rate should be $g^{\ast}$ (one can see this from the Hamiltonian, Eq.~(1)). In the analysis so far we have always taken the left and right coupling rates real and equal to each other. Now we shall focus our attention to the other interesting scenario where $g_{L} = g_{R}^{\ast}\neq g_R$, and we write  
\begin{equation}
g_{L}=|g|e^{i\Phi},\,\,\, 
g_{R}=|g|e^{-i\Phi}, \,\,\, {\rm with}\,\, \Phi\neq 0. 
\end{equation}
The angle $\Phi$ corresponds to the angular position of the atom.
We study how the two-photon quantum interference effects are modified when we vary $\Phi$. We will compare results for two nonzero values of $\Phi$, 
namely, $\Phi=\pi/8$ and $\Phi=\pi/4$, with the case we have treated so far, $\Phi=0$. 

Following the same calculations as before but with different angular positions we arrive at the results plotted in FIG.~5. The parameters have been chosen as in FIG.~2a. The most noticeable point is that as $\Phi$ is increased the red curve (corresponding to ab/ba detections) increases considerably. At $\Phi=\pi/4$ its maximum value approaches the maximum value reached for  the aa/bb detection density. Apparently, the amount of destructive interference
is the largest for $\Phi=0$, and then decreases to reach a minimum for $\Phi=\pi/4$ (not shown here is that for larger values of $\Phi$ the amount of destructive interference rises again).
\subsubsection{Early time behavior}
Before trying to explain this behavior we first emphasize that  there are many processes occurring (some nonlinear) for those later times when the maximum detection probability is reached. It does not seem possible to find understandable analytical expressions revealing the $\Phi$ dependence of those detection probabilities. In order, nonetheless, to gain some analytical understanding of the influence of the value of $\Phi$ we shall focus, instead, our attention to early times ($t\lesssim 1\kappa^{-1}$) when the dynamics in the weak coupling regime is relatively simple and interference effects can be understood more easily.
\begin{figure*}
\begin{center}
\begin{tabular}{cccc}
\subfloat{\includegraphics[width=8.5cm,height=6.85cm]{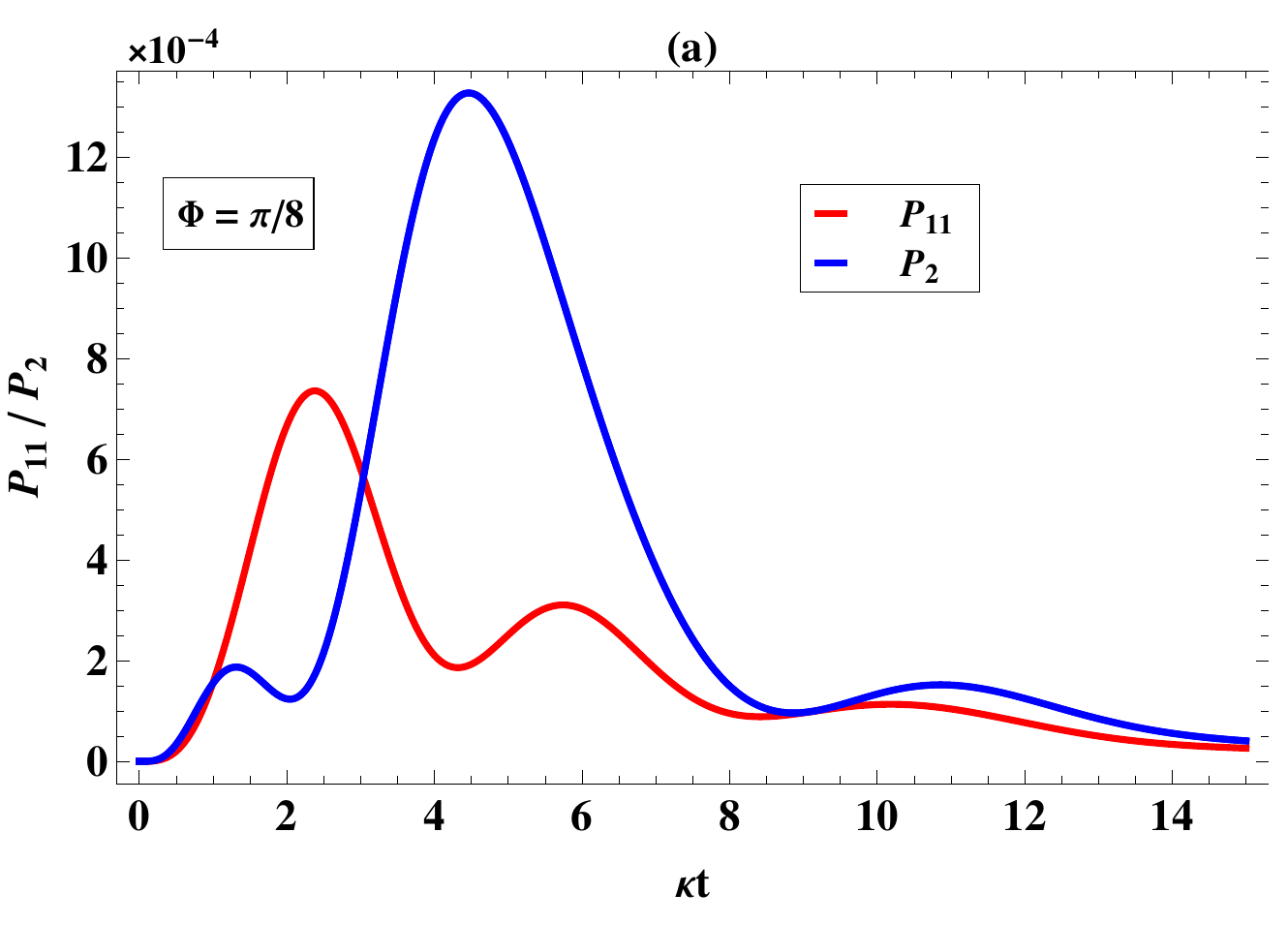}} & 
\subfloat{\includegraphics[width=9cm,height=7cm]{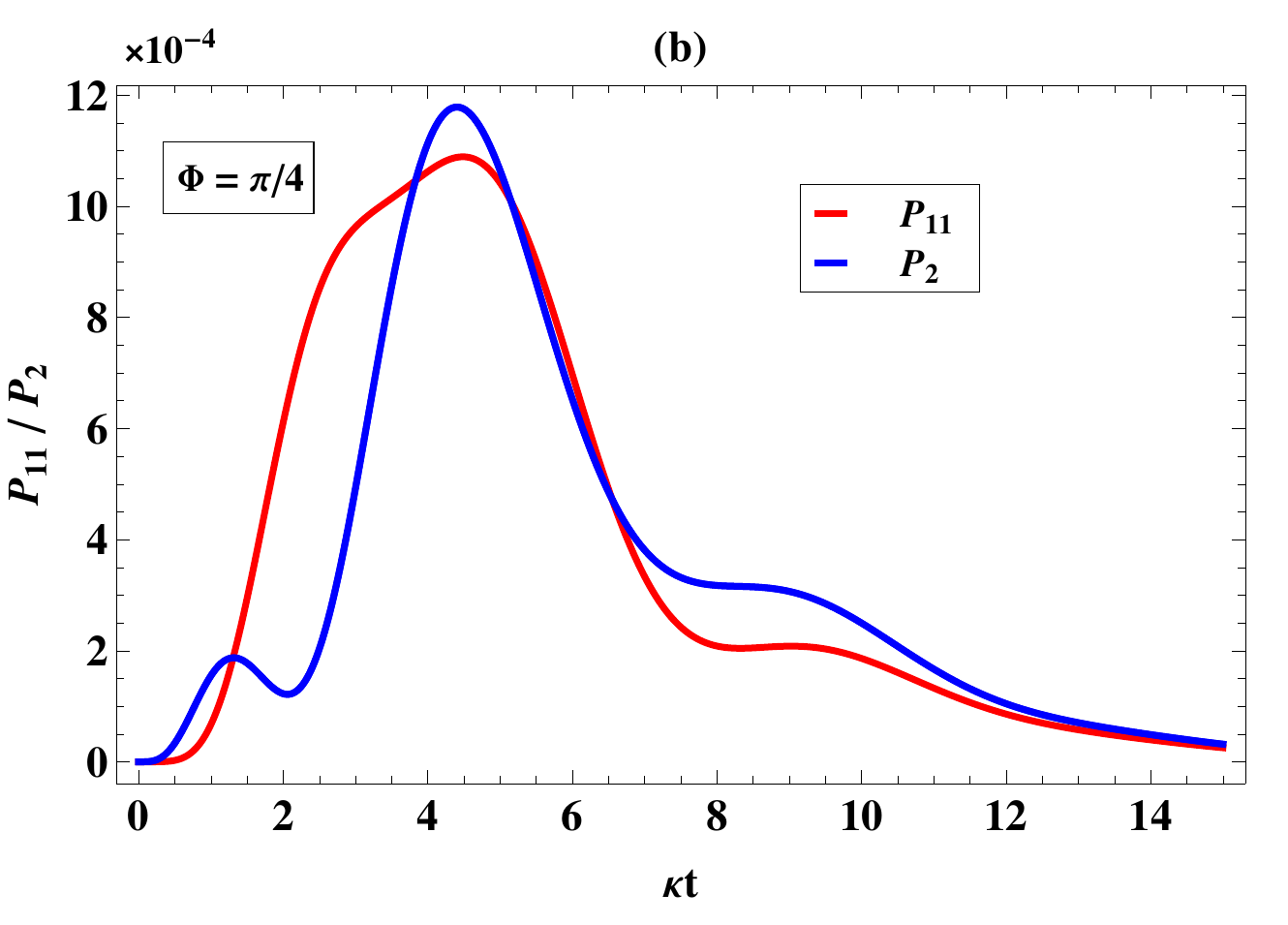}}\\
\end{tabular}
\captionsetup{
  format=plain,
  margin=1em,
  justification=raggedright,
  singlelinecheck=false
}
\caption{Behavior of the equal-time probability density of detecting photons at the output ports with varying angular position of the atoms (i.e., with varying phases $\Phi$ of the complex coupling rates). Two cases are shown: In (a) $\Phi=\pi/8$ and (b) $\Phi=\pi/4$.
In both plots the parameters are same as in Fig.[\ref{Fig2}-(a)], to which these plots should be compared. Note that with increasing $\Phi$ the red curve approaches the blue curve more and more. For initial times we try to understand this behavior in the next two figures.}\label{Fig5}
\end{center}
\end{figure*}

During these early times we can ignore the higher-order processes in which the photons leak out of one cavity and enter the other. The ``No-Jump state'' given in Eq.~[\ref{nojumptotal}] simplifies  considerably to
\begin{equation}\label{nojumpinitial}
\begin{split}
&\ket{\tilde{\psi}(t)}_{\kappa t\lesssim 1}=d_{1}(t)\ket{e_{1}00,e_{2}00}+d_{2}(t)\ket{e_{1}00,g_{2}10}\\
&+d_{3}(t)\ket{e_{1}00,g_{2}01}+d_{4}(t)\ket{g_{1}10,e_{2}00}+d_{5}(t)\ket{g_{1}01,e_{2}00}\\
&+d_{6}(t)\ket{g_{1}10,g_{2}10}+d_{7}(t)\ket{g_{1}10,g_{2}01}+d_{8}(t)\ket{g_{1}01,g_{2}10}\\
&+d_{9}(t)\ket{g_{1}01,g_{2}01}\\
&\equiv \Bigg(d_{11}(t)\ket{e_{1}00}+d_{12}\ket{g_{1}10}+d_{13}\ket{g_{1}01}\Bigg)\otimes\\
&\hspace{5mm} \Bigg(d_{21}(t)\ket{e_{2}00}+d_{22}\ket{g_{2}10}+d_{23}\ket{g_{2}01}\Bigg).
\end{split}
\end{equation}
The wave function simply factorizes into a left part and a right part, implying in particular that the left and right systems are not entangled with each other. The aa/bb and ab/ba equal-time detection probability densities now take the simple form
\begin{equation}\label{initialP2P11}
\begin{split}
&P_{2(\kappa t\lesssim 1)}=4\kappa^{2}\Bigg\vert d_{6}(t)\Bigg\vert^{2}{\rm \delta T},\\
&P_{11(\kappa t\lesssim 1)}=\kappa^{2}\Bigg\vert d_{7}(t)+d_{8}(t)\Bigg\vert^{2}{\rm \delta T}.
\end{split}
\end{equation}
FIG.~6 shows the time dependence of this lowest-order approximation to the  initial detection probability density for our three different values of $\Phi$. We note that for $\Phi=0$ both densities are almost the same. When $\Phi$ is increased from $\pi/8$ to $\pi/4$ the aa/bb probability dominates the ab/ba  probability. This indicates the by now familiar destructive interference mechanism working to decrease the chances of clicks at different detectors. We note that the double detection probability is hardly dependent on the value of $\Phi$. We can explain this by looking at the solutions of the coupled differential equations  obtained from the Non-Unitary Sch\"odinger equation. We note that these coupled differential equations describing the evolution of the probability amplitudes are homogeneous, linear and ordinary. We are mainly concerned with $d_{6}(t), d_{7}(t)$ and $d_{8}(t)$ as only these amplitudes contribute to the detection densities. The full analytical solution of this set of coupled differential equations is rather involved and won't be displayed here, but the solutions consist of sums of exponential functions, with complex exponents $\lambda_{i}$) multiplied by complex amplitudes $\alpha_{i}$ (for $1\leq i \leq 5$). 
It turns out then that for $\Phi=0$ all three amplitudes are almost equal which explains why we were getting almost the same probability densities for both detection types for this case. For $\Phi=\pi/4$ the amplitudes $d_{6}(t)$, $d_{7}(t)$ and $d_{8}(t)$ have different values for the exponents $\lambda_{i}$ and for the amplitudes $\alpha_{i}$. There are certain terms in $d_{7}(t)$ and $d_{8}(t)$ that have almost the same magnitude but with opposite phases. This is causing the destructive interference behavior visible in the plots. 
\begin{figure*}
\begin{center}
\begin{tabular}{cccc}
\subfloat{\includegraphics[width=6cm,height=5cm]{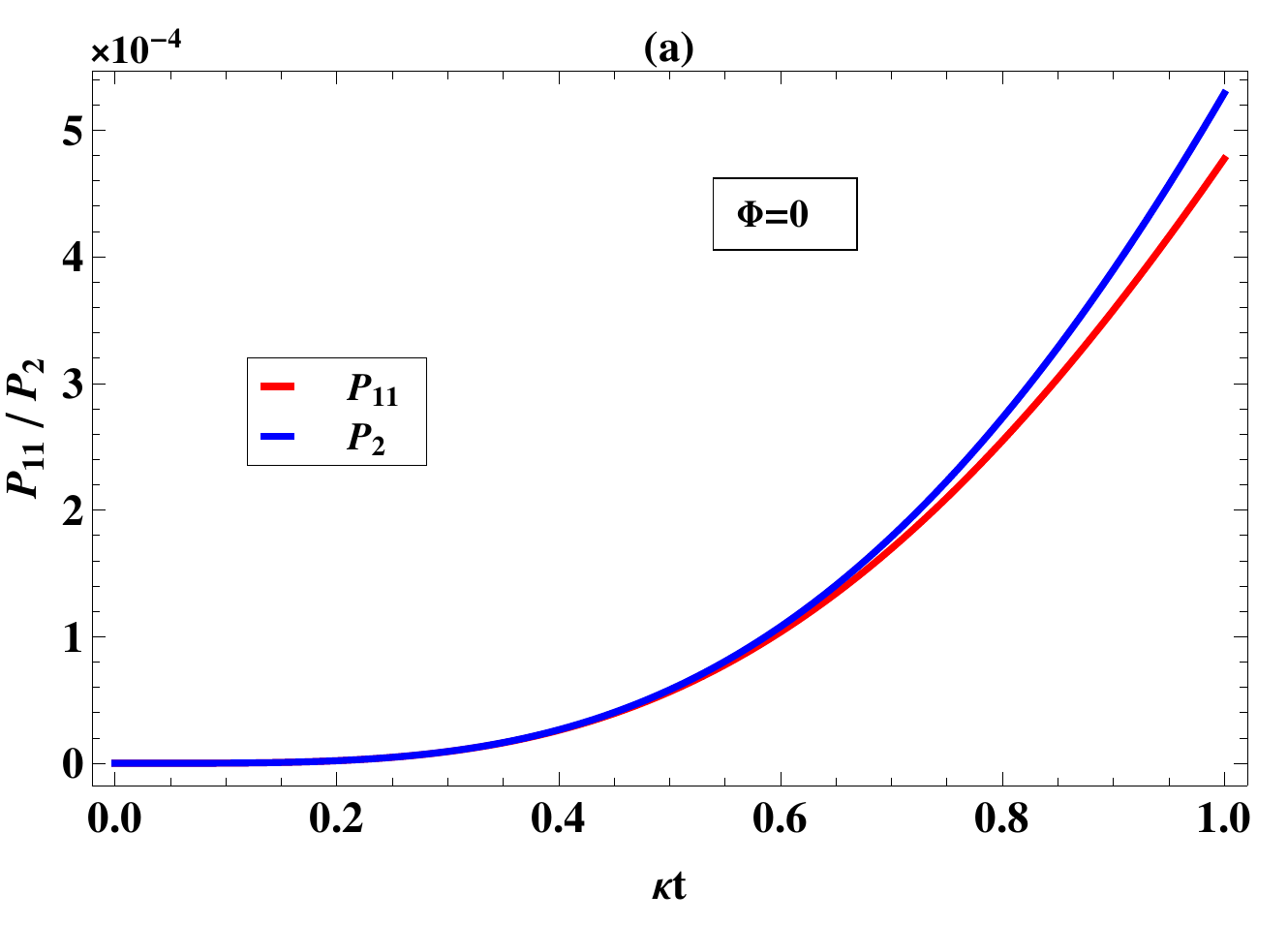}} & 
\subfloat{\includegraphics[width=6cm,height=5cm]{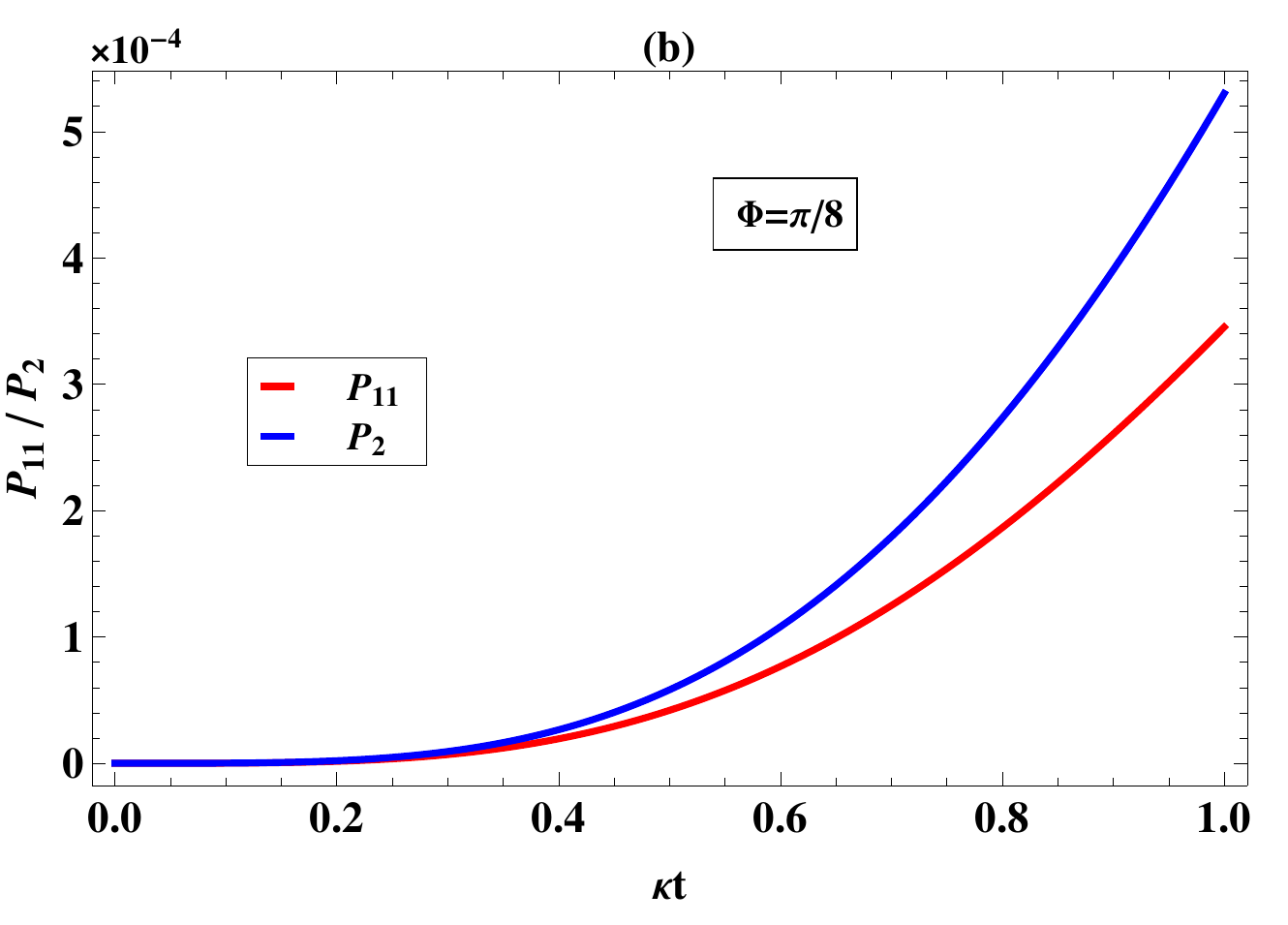}}&
\subfloat{\includegraphics[width=6cm,height=5cm]{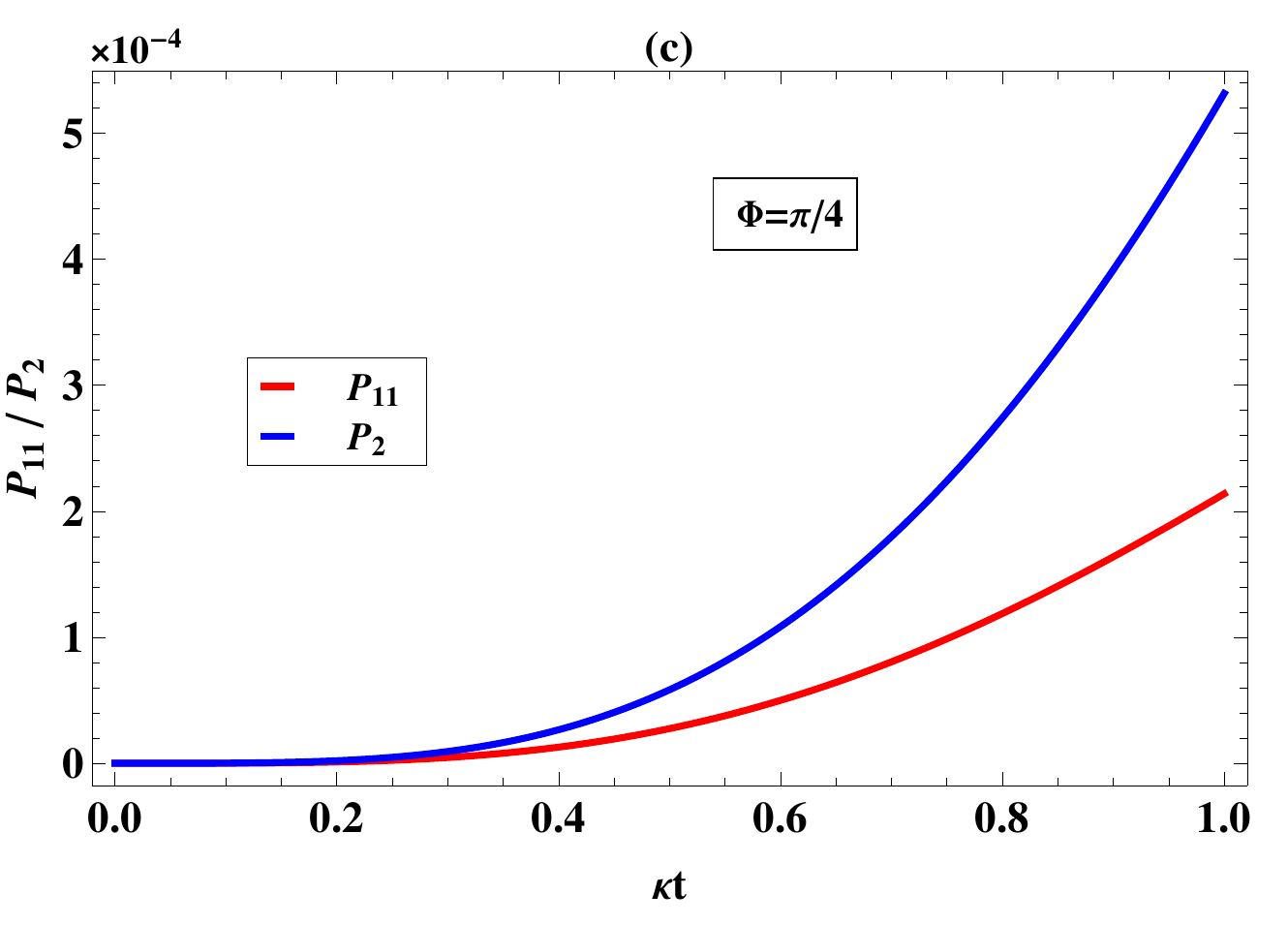}}\\
\end{tabular}
\captionsetup{
  format=plain,
  margin=1em,
  justification=raggedright,
  singlelinecheck=false
}
\caption{Temporal profile of the equal-time detection probability densities for initial times ($t\lesssim 1\kappa^{-1}$) for different values  of $\Phi$. Parameters are the same as in FIG.~[2(a)]. The plots shows a decrease in the ab/ba density (red curves) with increasing $\Phi$ while the aa/bb density (blue curves) maintains its value for all three values of $\Phi$.}\label{Fig6}
\end{center}
\end{figure*}

\begin{figure}
\begin{center} 
\subfloat{\includegraphics[width=8cm,height=6.4cm]{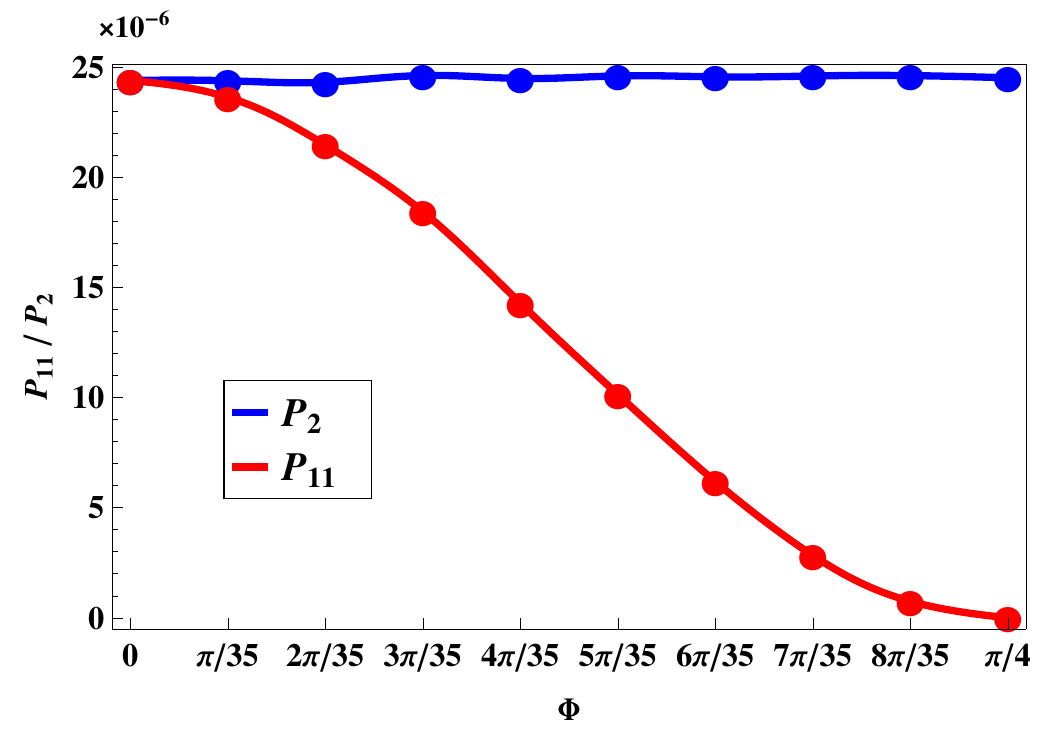}}
\captionsetup{
  format=plain,
  margin=1em,
  justification=raggedright,
  singlelinecheck=false
}
\caption{At a fixed time, $t=0.4\kappa^{-1}$, we plot the equal-time ab/ba and aa/bb detection probability densities versus $\Phi$ between the range $0$ and $\pi/4$. The aa/bb detection probability remains constant, whereas the ab/ba detection probability decays as a function of phase. Parameters are the same as in FIG.~[2(a)].}\label{Fig7}
\end{center}
\end{figure}

Moreover, we analyzed the effect of changing the phase $\Phi$ continuously between $0$ and $\pi/4$ at some fixed early time.  For small times we can simplify the exact numerical solution by Taylor expanding the exponentials around $t=0$. Because of the initial conditions we have $d_{1}(t=0)=1$ and all other amplitudes are initially zero.
The zeroth-order term in the series expansion of $d_{6}(t)$, $d_{7}(t)$ and $d_{8}(t)$ is, therefore, zero. The first-order terms of detection probabilities also vanish (this is to be expected as our probability densities refer to {\em two} (joint) detection events), and hence we have to keep the terms till second order in time. Then, in the notation introduced above, these second-order terms turn out to be
\begin{equation}
\begin{split}
& d_{6}(t)\cong\sum_{i=1}^{5}\frac{\alpha^{(6)}_{i}\lambda^{(6)^{2}}_{i}}{2}t^{2},\\
& d_{7}(t)+d_{8}(t)\cong\sum_{i=1}^{5}\Bigg[\frac{(\alpha^{(7)}_{i}+\alpha^{(8)}_{i})\lambda^{(7)^{2}}_{i}}{2}\Bigg]t^{2}.
\end{split}
\end{equation}
Superscripts $(k)$ for $k=6,7,8$ were introduced here to distinguish among exponents and complex amplitudes belonging to different amplitudes $d_k$. Note that the exponents appearing in $d_{7}(t)$ and $d_{8}(t)$ are the same and hence they are given the same name of $\lambda^{(7)}_{i}$. In FIG.~7 we plot the detection probability densities (the absolute value squared of the quantities defined in the last equation) versus $\Phi$ between $0$ and $\pi/4$. 
We have chosen a fixed time $t=0.4 \kappa^{-1}$ and find that with increasing $\Phi$ the ab/ba probability density decreases due to destructively interfering terms while the aa/bb detection probability remains practically constant.

The plots in FIG.~6 and FIG.~7 are consistent with the initial part of FIG.~5. At the end of this subsection we once again emphasize that for later times many additional (higher-order) processes occur. But by looking at the joint detection probabilities for initial times we have at least shown and understood that interference effects do play an important role, and that those effects do depend on the angular positions of the atoms through $\Phi$.

\section{Entanglement}
In this Section we  quantify the amount of bipartite entanglement generated in our system from the initial unentangled state. We first consider the case where both excitations are still in the system (i.e., before any photon has been detected), and then we consider the entanglement left after one photon has been detected (of course, once both photons have been detected no entanglement remains).

There are several different possible measures of entanglement that can be used to quantify bipartite entanglement \cite{guhne2009entanglement,von1995mathematische}.  One measure, the negativity, follows from the Peres-Horodecki Positive Partial Transpose (PPT) criterion \cite{horodecki1996separability,peres1996separability,kurzyk2012introduction}, and it can be calculated for systems of arbitrary Hilbert space dimensions \cite{VidalWerner}. This quantity is appropriate for quantifying the entanglement between the left and right systems (atom plus cavity modes) as a whole. If we are interested in the entanglement between just the atoms, we may also calculate the concurrence. If the state of our system happens to be pure, we may in addition calculate the Von Neumann entropy of the reduced system and use it to quantify entanglement.

\subsection{Two excitations}
\begin{figure*}[t]
\begin{center}
\begin{tabular}{cccc}
\subfloat{\includegraphics[width=8.8cm,height=6.5cm]{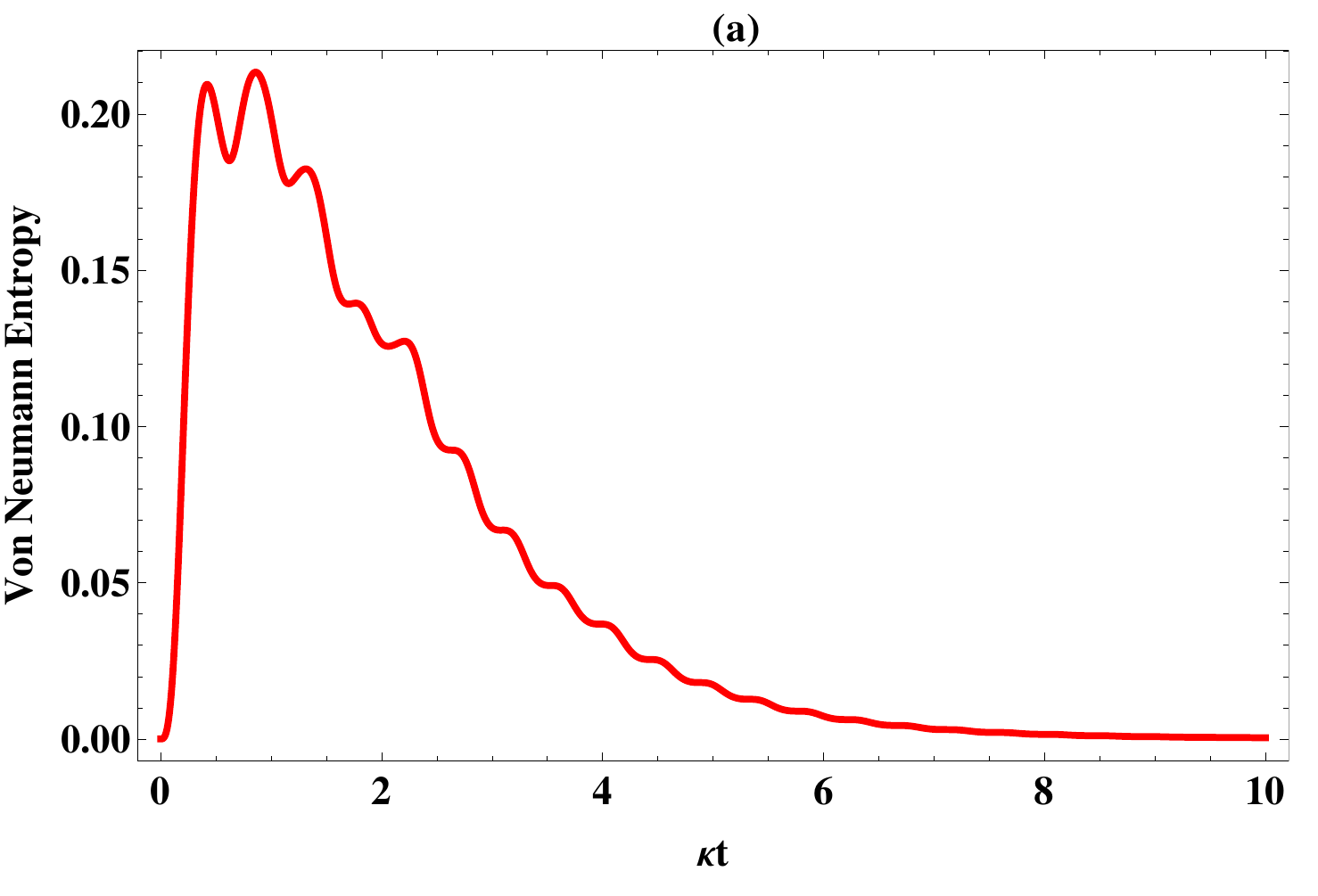}} & 
\subfloat{\includegraphics[width=8.8cm,height=6.5cm]{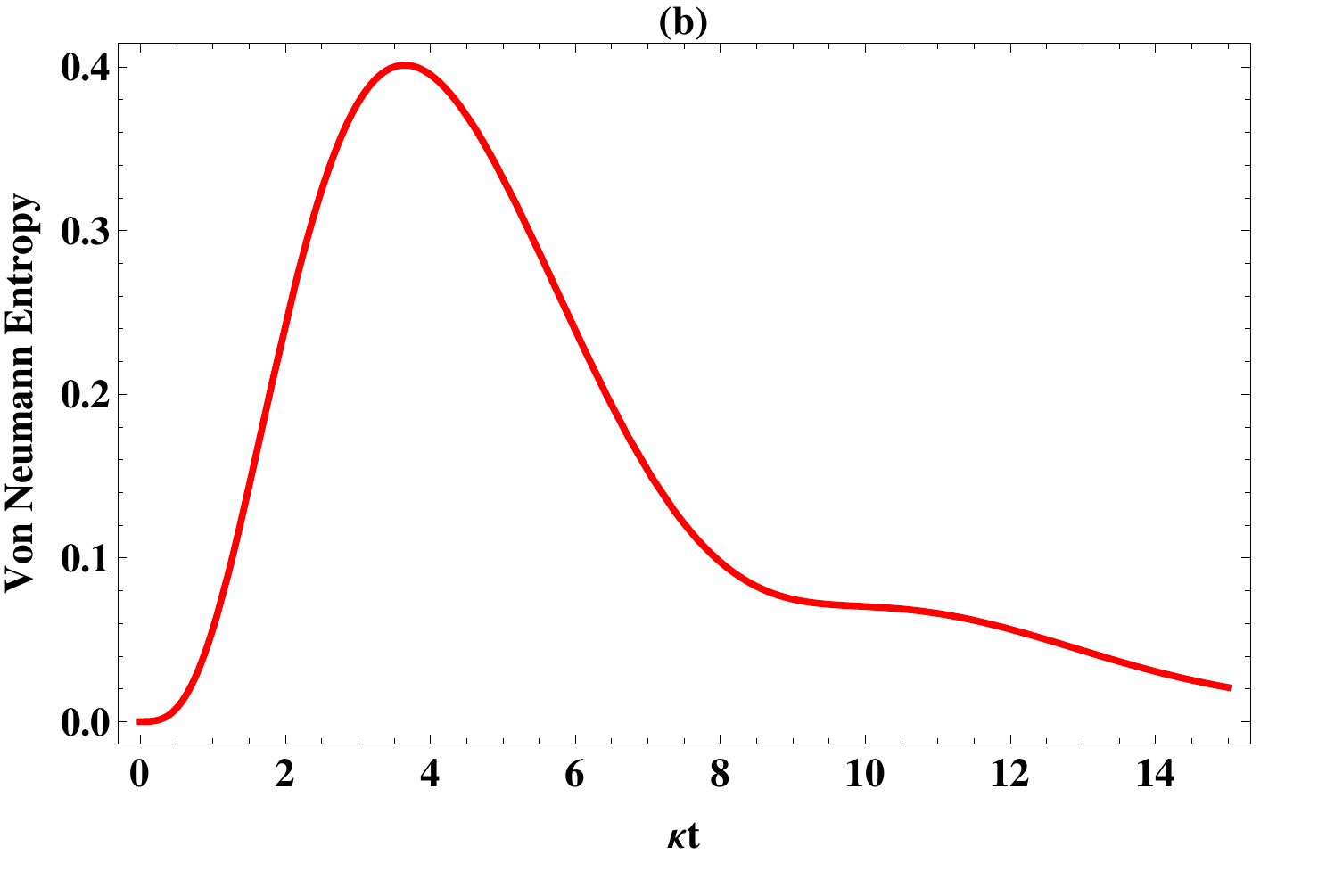}}\\
\end{tabular}
\captionsetup{
  format=plain,
  margin=1em,
  justification=raggedright,
  singlelinecheck=false
}
\caption{(a) Von Neumann entropy as a measure of entanglement in (a) strong coupling regime with parameters $|g|/\kappa=5, \Delta/\kappa=0.5$ and (b) in weak coupling regime with all other parameters same as in part (a) except $|g|/\kappa=0.25$. }\label{Fig8}
\end{center}
\end{figure*}

Here we first calculate the amount of entanglement when no excitation is lost by the system, i.e. for the no jump situation. This calculation is simple because the state of the system before photon detection is pure.
In the case of a pure state $\ket{\psi(t)}$ of two subsystems the (binary, base-2) Von Neumann Entropy $E(\ket{\psi(t)})$ \cite{schneiderbauer2012entanglement,audretsch2008entangled,wilde2013quantum} is an appropriate and physically intuitive measure of entanglement. It is defined as
\begin{equation}
E\Bigg(\ket{\tilde{\psi}(t)}_{LR}\Bigg)= -{\rm Tr}[\rho_{L}{\rm log_{2}}\rho_{L}]=-{\rm Tr}[\rho_{R}{\rm log_{2}}\rho_{R}],
\end{equation}
which is evaluated by means of the eigenvalues $\lambda_{i}$ of $\rho_{L}$ (or, equivalently, of $\rho_{R}$) as
\begin{equation}
E\Bigg(\ket{\tilde{\psi}(t)}_{LR}\Bigg)= -\sum_{i}\lambda_{i}{\rm log_{2}}\lambda_{i}.
\end{equation}
The curves in FIG.~8 give the plots of Von Neumann entanglement in both the strong ($|g|>\kappa$, plot (a)) and the weak coupling regimes ($|g|<\kappa$, plot (b)). In the strong coupling regime the presence of Rabi oscillations is manifested by oscillations in the amount of entanglement. The oscillations have a small amplitude (in the next Figure we will encounter large-amplitude oscillations) here, because the entanglement between left and right systems does not change by just the exchange of an excitation between an atom and one of the cavity modes.
Entanglement in the weak coupling regime is almost twice as large as in the strong coupling case, because entanglement is caused by photons travelling back and forth between the cavities. Due to the presence of dissipation (cavity decay), the entanglement never becomes maximal and reaches at most a value of about 0.4 ebits (an ebit is the standard unit of Von Neumann entanglement, corresponding to a maximally entangled state of two qubits). 

\subsection{One excitation}
In this subsection we quantify entanglement left in the system after one photon has been detected. We will use both concurrence and negativity as measures of entanglement.

We first write down the general form of the  state of our system in the case that a single excitation is left. We'll use the same notation convention as before, but with the probability amplitudes  now denoted by $f_{i}(t)$ with $1\leq i \leq6$, as follows:
\begin{equation}\label{singlenojump}
\begin{split}
&\ket{\tilde{\psi}(t)}=f_{1}(t)\ket{e_{1},0,0,g_{2},0,0}+f_{2}(t)\ket{g_{1},0,0,e_{2},0,0}\\
&+f_{3}(t)\ket{g_{1},1,0,g_{2},0,0}+f_{4}(t)\ket{g_{1},0,1,g_{2},0,0}\\
&+f_{5}(t)\ket{g_{1},0,0,g_{2},1,0}+f_{6}(t)\ket{g_{1},0,0,g_{2},0,1}.
\end{split}
\end{equation}
\subsubsection{Concurrence between atoms}
The two atoms in our system constitute a pair of qubits, which are coupled to their respective cavities. The state of the two atoms is, generally, mixed (not pure). For this specific case of two qubits forming a mixed state the concurrence $C(\rho)$ as first introduced by Wootters \cite{hill1997entanglement,wootters1998entanglement} is an appropriate measure of entanglement. 
We first construct the atomic density operator $\hat{\rho}_{a}$ from the total density matrix by taking the trace over all four cavity modes. This yields: 
\begin{equation}\label{rhoA}
\begin{split}
&\hat{\rho}_{a}={\rm Tr}_{{\rm cav.}}[\hat{\rho}(t)]\\
&\hspace{4mm}=|f_{1}(t)|^{2} \ket{e_{1},g_{2}}\bra{e_{1},g_{2}}+|f_{2}(t)|^{2}\ket{g_{1},e_{2}}\bra{g_{1},e_{2}}+\\
&\hspace{5mm} f_{1}(t)f_{2}^{\ast}(t)\ket{e_{1},g_{2}}\bra{g_{1},e_{2}}+f_{2}(t)f_{1}^{\ast}(t)\ket{g_{1},e_{2}}\bra{e_{1},g_{2}}\\
&\hspace{4mm} +(1-|f_{1}(t)|^{2}-|f_{2}(t)|^{2})\ket{g_{1},g_{2}}\bra{g_{1},g_{2}}.
\end{split}
\end{equation}
Following Wootters  we can express the concurrence $C(t)$ as
\begin{equation}\label{C}
C(t)=\max\Bigg(0,\sqrt{\lambda_{1}}-\sqrt{\lambda_{2}}-\sqrt{\lambda_{3}}-\sqrt{\lambda_{4}}\Bigg),
\end{equation} 
where $\lambda_{i}$'s are the eigenvalues (in descending order of magnitude) of the spin flipped density matrix $\widetilde{\rho}=\hat{\rho}_{a}(\hat{\sigma}_{y}\otimes\hat{\sigma}_{y})\hat{\rho}^{\ast}_{a}(\hat{\sigma}_{y}\otimes\hat{\sigma}_{y})$, with $\hat{\sigma}_{y}$ being the Pauli spin flip operator. Using Eq.~[\ref{rhoA}] in Eq.~[\ref{C}] we arrive at the following simple expression for the atomic concurrence,
\begin{equation}
C(t)=2|f_{1}(t)||f_{2}(t)|.
\end{equation}
It  can take any value between 0 to 1, where 0 refers to a completely separable state and 1 to a maximally entangled (pure) state.

In FIGs.~9 (a) and (c) we have plotted the atomic concurrence in both the strong and weak coupling regimes, respectively. Our results are consistent with previously reported results in \cite{montenegro2012entanglement,casagrande2008dynamics}. In both regimes, we have chosen parameters such that the photon should remain trapped in the cavities for longer times so that entanglement can be sustained for long enough times. FIG~9(b) is plotted to show this explicitly by varying the cavity decay rate $\kappa$ in the weak coupling regime. We notice that the concurrence achieves its maximum value after a time on the order of one or two $\kappa^{-1}$, simply because bipartite entanglement is created by photons leaking out of one cavity and travelling to the other, which takes a time on the order of $\kappa^{-1}$. We also note that in the strong coupling regime the concurrence displays oscillatory behavior which originates from the single-photon Rabi oscillations (between atom and cavity mode(s)), while in the weak coupling regime the concurrence displays an almost purely decaying behavior. In contrast to the situation displayed in FIG.~8, here the Rabi oscillations may make the entanglement between atoms completely disappear, namely, when the excitation is transferred to a cavity mode.
After long enough times (in the limit $\kappa t>>1$ in the strong coupling and $gt>>1$ in the weak coupling case) the atoms again become unentangled once they end up in their ground states.

\begin{figure*}
\begin{center}
\begin{tabular}{cccc}
\subfloat{\includegraphics[width=5cm,height=4.5cm]{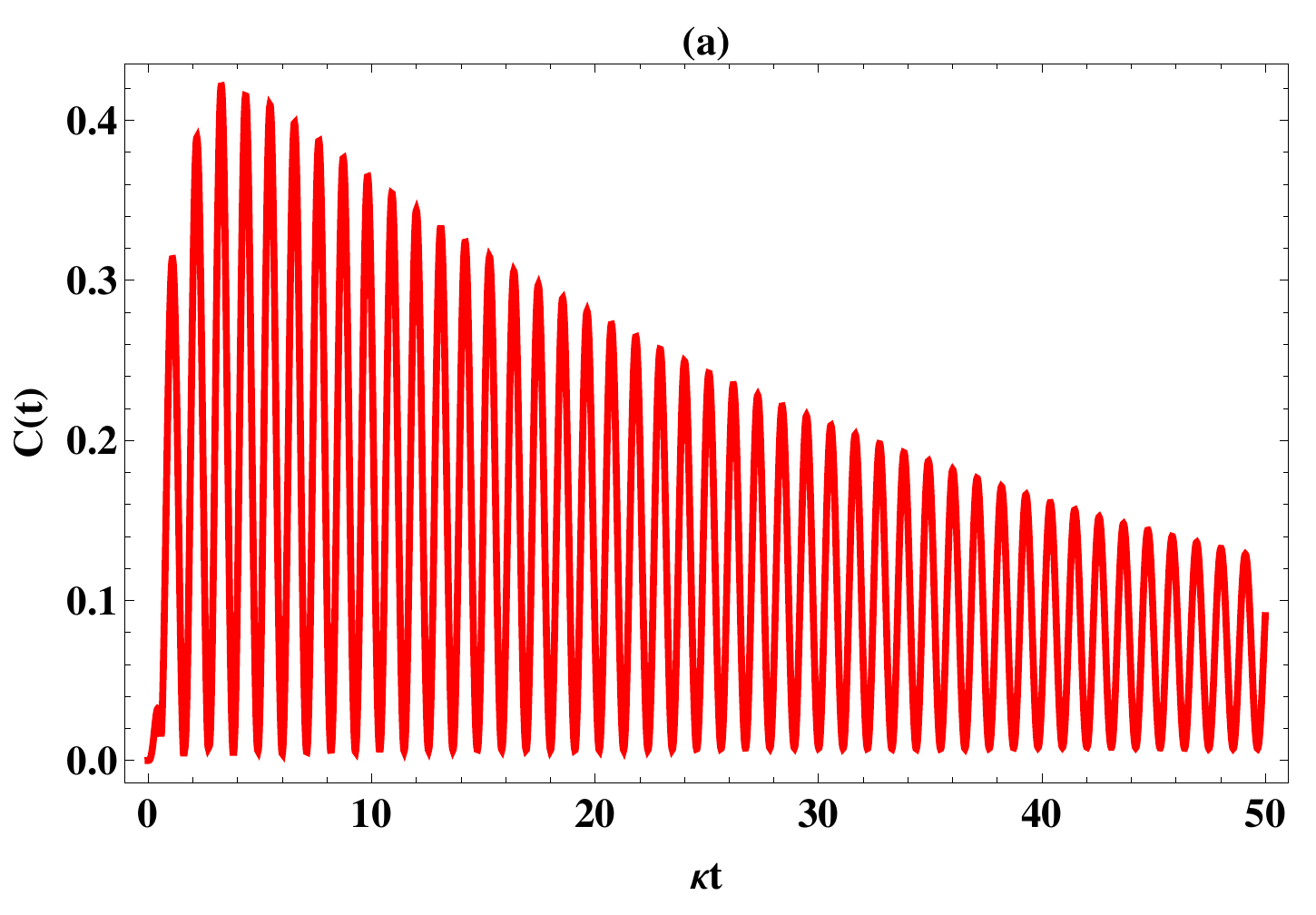}} & 
\subfloat{\includegraphics[width=6.5cm,height=4.5cm]{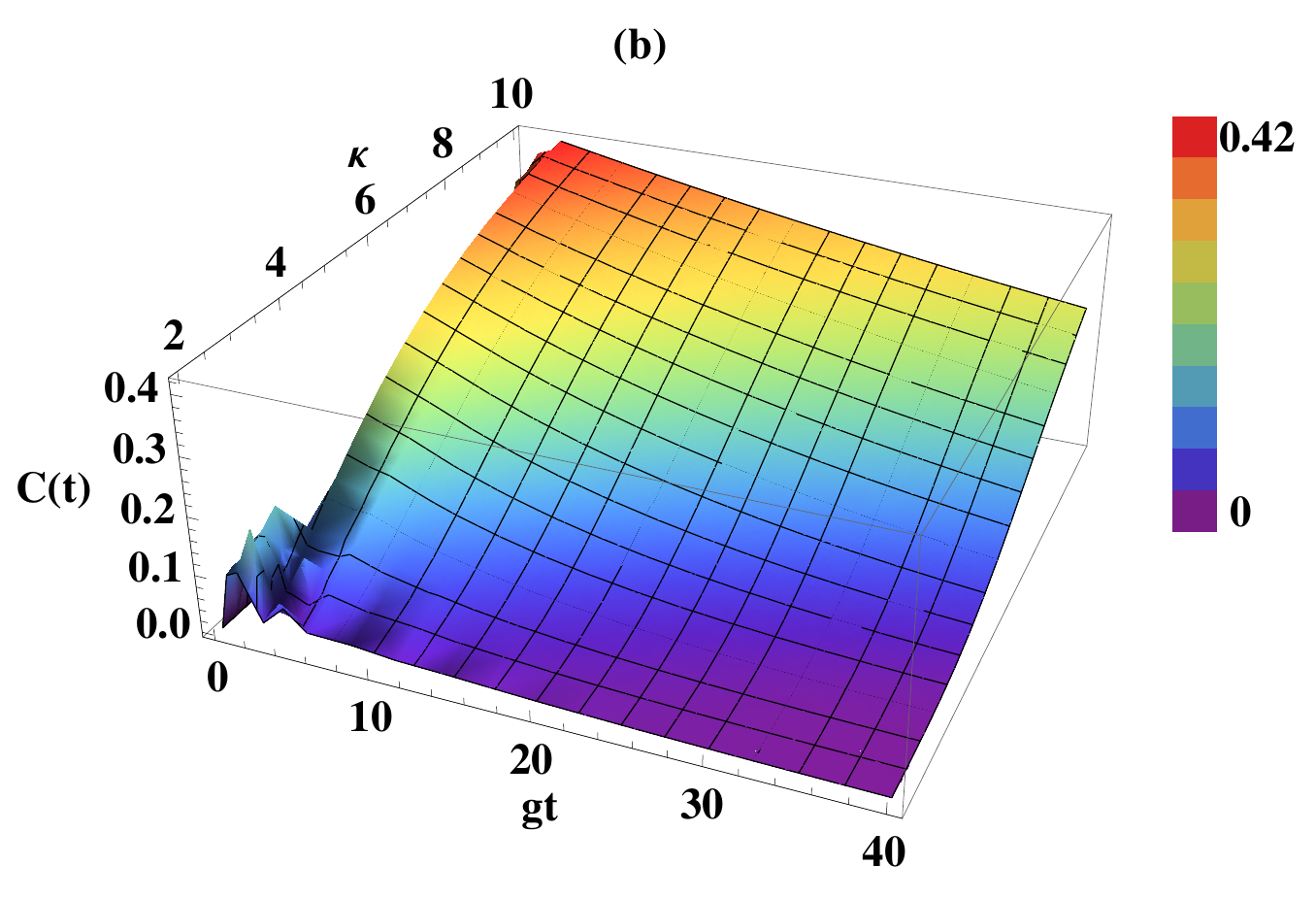}} & 
\subfloat{\includegraphics[width=5cm,height=4.5cm]{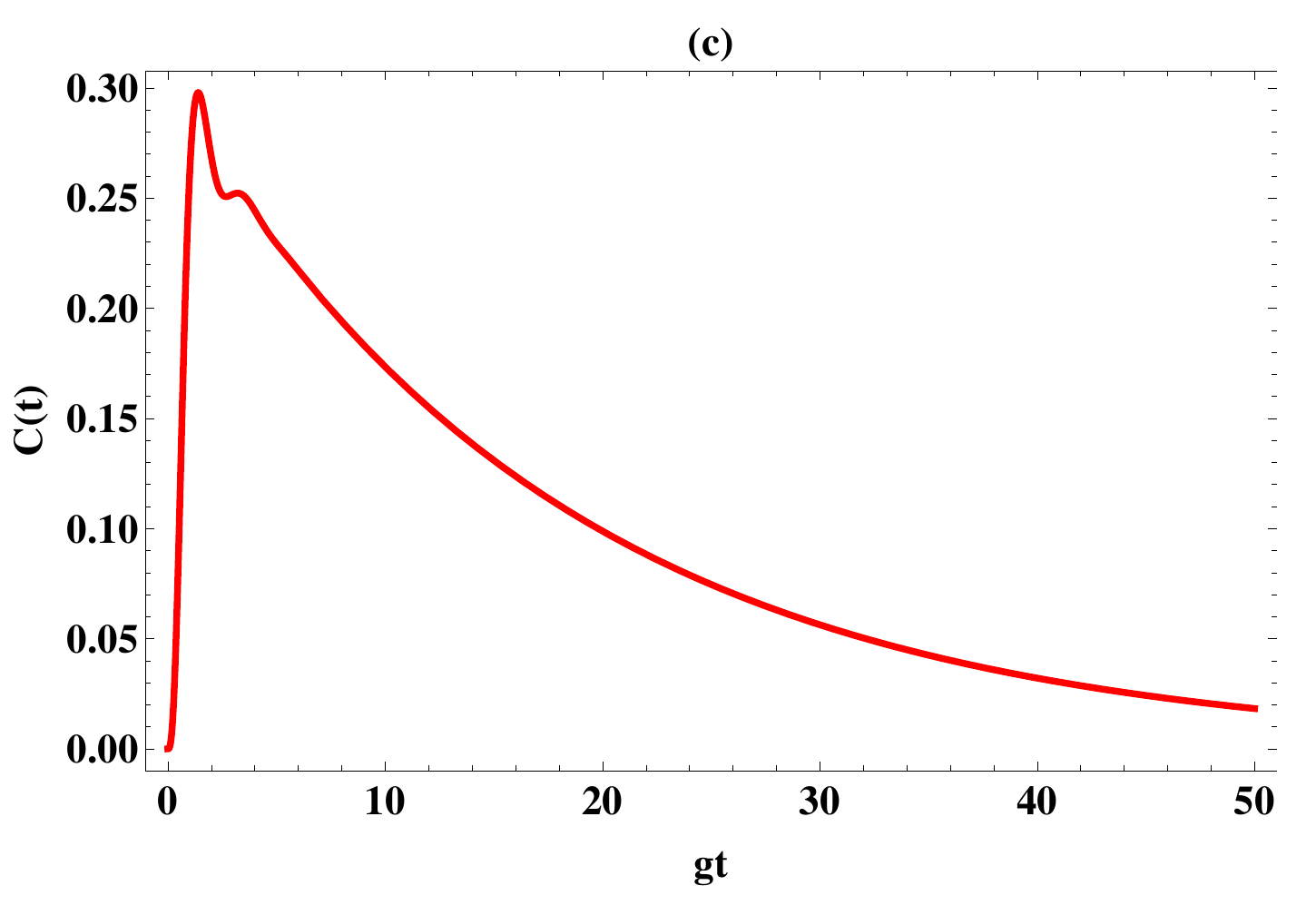}}\\
\end{tabular}
\captionsetup{
  format=plain,
  margin=1em,
  justification=raggedright,
  singlelinecheck=false
}
\caption{Time evolution of the concurrence between the two atoms. (a) Strong coupling regime with parameters $|g|/\kappa=2, \Delta_{c}/\kappa=0.5$. (b) Concurrence as a function of time and $\kappa$ (which varies from 2$|g|$ to 10$|g|$ so as to stay in the weak coupling regime). Note that the concurrence increases with increasing $\kappa$. (c) Weak coupling regime with parameters $\kappa/|g|=5$, $ \Delta_{c}/|g|=0.5$.}\label{Fig9}
\end{center}
\end{figure*}

\subsubsection{Negativity between two atoms}
The negativity $N$ is a more flexible quantity, as it can be (easily) used to quantify entanglement of systems with larger Hilbert spaces \cite{VidalWerner}. It is defined by
\begin{equation}\label{Nega}
N=\max\Bigg(0,-2\sum_{i}\lambda_{i}\Bigg)
\end{equation}
where the sum is taken over the negative eigenvalues $\lambda_{i}$ of the partially transposed atomic density matrix $\hat{\rho}_{a}$ (given in Eq.[\ref{rhoA}]). Partial transposition is taken with respect to one of the atoms only and here we'll perform it with respect to atom B. Like the concurrence, $N$ ranges between 0 and 1.
The partially transposed atomic density matrix (with respect to atom B) $\rho_{a}^{(B)}$ is given by:
\begin{equation}
\begin{split}
&\hat{\rho}_{a}^{(B)}=|f_{1}(t)|^2\ket{e_{1},g_{2}}\bra{e_{1},g_{2}}+|f_{2}(t)|^2\ket{g_{1},e_{2}}\bra{g_{1},e_{2}}+\\
& f_{1}(t)f_{2}^{\ast}(t)\ket{e_{1},e_{2}}\bra{g_{1},g_{2}}+f_{2}(t)f_{1}^{\ast}(t)\ket{g_{1},g_{2}}\bra{e_{1},e_{2}}+\\
& (1-|f_{1}(t)|^2-|f_{2}(t)|^2)\ket{g_{1},g_{2}}\bra{g_{1},g_{2}}.
\end{split}
\end{equation}
In the next step we express the partially transposed density operator in matrix form, and for that we used the basis set $\lbrace\ket{g_{1},g_{2}},\ket{e_{1},g_{2}},\ket{g_{1},e_{2}},\ket{e_{1},e_{2}}\rbrace$, noticing that even though our original problem is limited to a single excitation, after taking the partial transpose we have to include one state with two excitations in the basis as well. After calculating the negative eigenvalues of the matrix $\hat{\rho}_{a}^{(B)}$, we plug those values into Eq.[\ref{Nega}] to get the desired negativity between the atoms. 
\begin{figure*}
\begin{center}
\begin{tabular}{cccc}
\subfloat{\includegraphics[width=9cm,height=7cm]{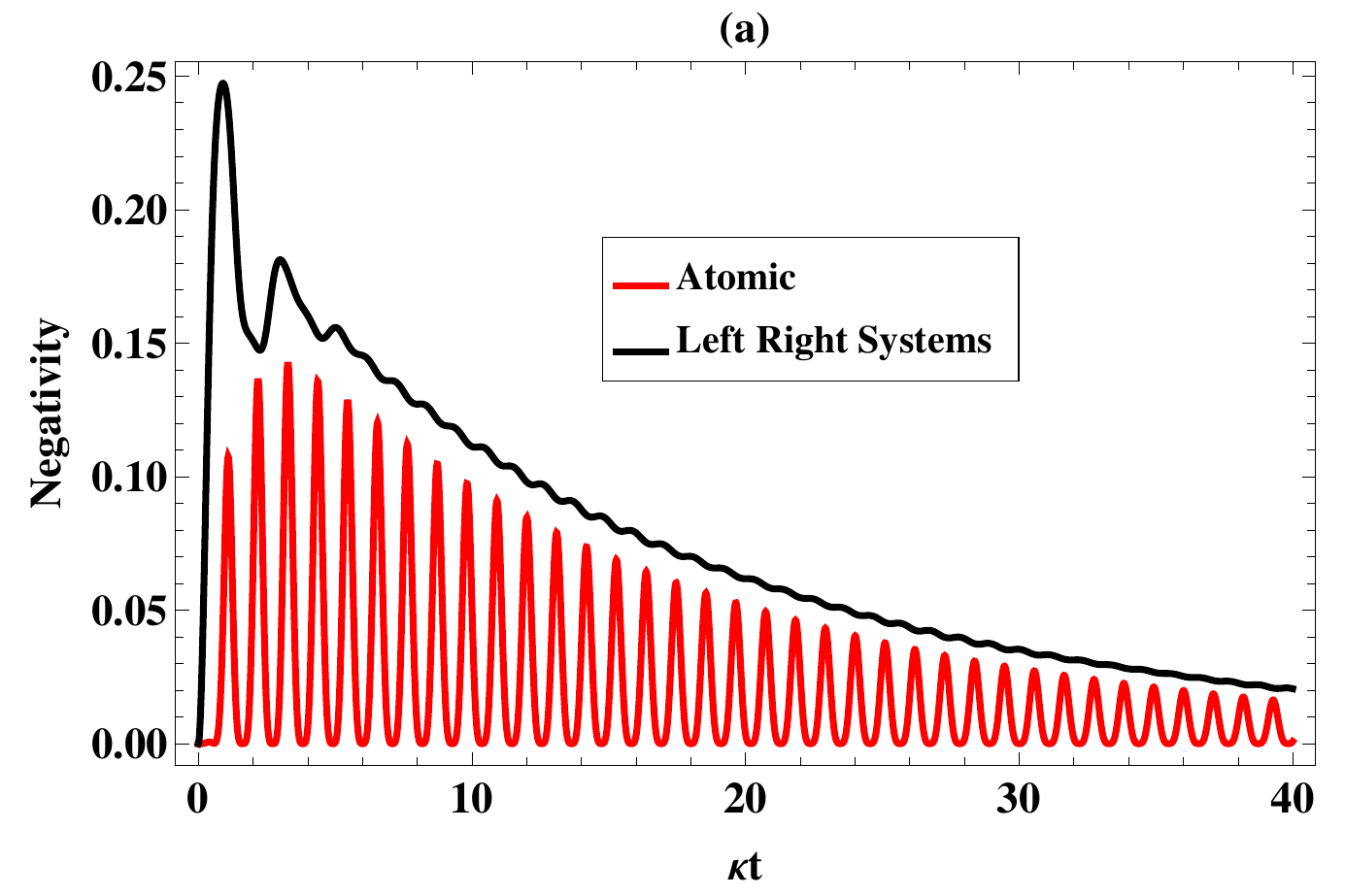}} & 
\subfloat{\includegraphics[width=9cm,height=7cm]{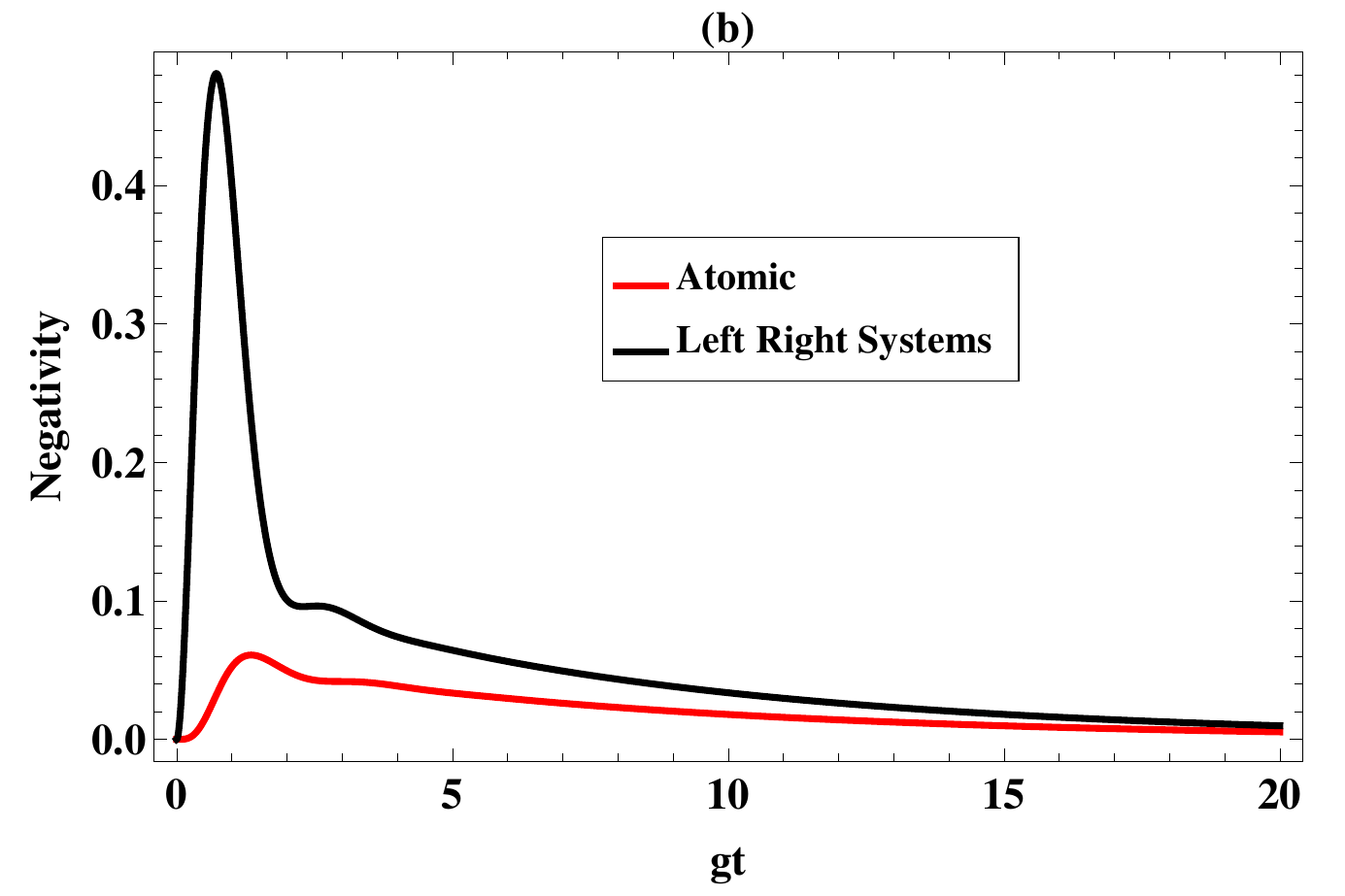}}\\
\end{tabular}
\captionsetup{
  format=plain,
  margin=1em,
  justification=raggedright,
  singlelinecheck=false
}
\caption{Dynamics of the negativity between two atoms (red curves) and between left and right systems (black curves) in (a) strong and (b) in weak coupling regime for the parameters same as in FIG.~9. Like for the concurrence, in the strong coupling regime the negativity's maximum value is greater than for the weak coupling regime. Also, negativity time evolution follows same behaviour as of concurrence but the maximum value achieved by negativity values are much smaller in both regimes, compared to concurrence (see FIG.~9 for comparison). As expected negativity in the case of left and right systems is considerably greater than the negativity between two atoms and like before there are oscillations in strong coupling regime and pure decay in weak coupling case.
}
\end{center}
\end{figure*}
In FIG.~10 (red curves) we have plotted the atomic negativity in the strong and weak coupling regimes. In comparison to the concurrence plots with the same parameters, we note that the overall shape of the temporal profile looks very similar, but the maximum values are much smaller in both regimes. This is consistent with a general theorem which compares the negativity and the concurrence in the case of mixed states of two qubits, which is discussed in greater detail in \cite{verstraete2001comparison,eisert1999comparison}.

\subsubsection{Negativity between left and right systems}
As long as the single excitation shuttles from one cavity to the other we can think of the left and right parts (including atoms {\em and} cavity modes) of the system as being entangled. For this specific calculation the negativity is about the only computable measure of entanglement we can use \cite{VidalWerner}. Following the procedure introduced in last subsection we calculated the negativity between system L and R. The main difference is that now we have to include more basis states to express the partially transposed density operator in matrix form. FIG.~10 (black curves) shows the corresponding plots of the negativity.

There are several noticeable points here. Firstly, the total negativity ($N_{LR}$) poses an upper bound on the atomic negativity ($N_{a}$), simply because the atoms are subsystems of systems L and R. Secondly, the entanglement between the L and R systems reaches its maximum earlier than does the atomic entanglement, indicating that the atoms become entangled only after the cavity modes become entangled with each other. This agrees with the picture that it is the photons travelling between cavities that generate the entanglement, and the travelling photons have to enter the cavity before they can (re)excite the atom.
Thirdly, even choosing the parameters such that the photon remains trapped in the cavities for longer times leads only to at most about 0.46 units of entanglement, due to the presence of various decay mechanisms in our system. Effects of loss mechanisms on the maximum value of negativity for multiqubit systems  (for both Markovian and non-Markovian baths) is discussed in great detail in \cite{man2010entanglement}, and our results  here are consistent with the conclusions reached in that article.
Fourthly, in the strong coupling regime one sees the usual Rabi oscillations in the atomic entanglement, but not in the system entanglement. The reason is that the local transfer of an excitation between atom and cavity mode does not affect the latter type of entanglement, but it does affect the former type of entanglement.

\section{Conclusions}
We studied in some detail interference effects between two photons in a coupled cavity system, as well as bipartite entanglement between two atom-cavity systems, as mediated by the photons traveling back and forth between the two systems. 

The interference is of the Hong-Ou-Mandel type, and our calculations showed that the two photons are in general more likely to be detected by one and the same detector, rather than by two different detectors.
The destructive interference between different pathways leading to the photons ending up in different detectors thus survives both nonlinear optics effects (due to the presence of atoms in our cavities) and spectral filtering by the resonant cavities. 

Our quantitative calculations of entanglement confirm several intuitive properties of our system and the entanglement therein: First, it takes one or two cavity decay times to build up entanglement, because it is mediated by the photons traveling between the two atom-cavity systems. Second, by considering entanglement between just the atoms on the one hand, and between the atom-cavity systems on the other, we confirm that the latter provides an upper bound on the amount of entanglement between the atoms. Moreover, whereas entanglement between the atom-cavity systems is not affected  by the Rabi oscilations of the excitation between the atom and the cavity it is in, the entanglement between the atoms does disappear when  excitations are transferred to the cavity mode.

\section*{Acknowledgements}
The authors are grateful to Jeff Kimble for originating some of the ideas presented here, and for useful comments on the manuscript. 
\bibliography{ECC3}
\end{document}